\documentclass[fleqn,usenatbib,useAMS]{mnras}

\usepackage{ae,aecompl}
\usepackage{bm}
\usepackage[utf8]{inputenc}
\usepackage{upgreek}
\usepackage{ulem}
\usepackage{amsmath}
\usepackage{amsfonts}
\usepackage{amssymb}
\usepackage{appendix}
\usepackage{graphicx}
\usepackage[dvipsnames]{xcolor}
\usepackage{cancel}
\newcommand{\od}{\mathrm{d}}        
\newcommand{\dd}{\mathrm{d}}        
\newcommand\deriv[2]{\frac{\partial#1}{\partial#2}}
\newcommand\oderiv[2]{\frac{\od#1}{\od#2}}
\newcommand\dderiv[2]{\frac{\text{D}#1}{\text{D}#2}}
\newcommand\mean[1]{\overline#1}
\newcommand\meanh[1]{\langle#1\rangle_{xy}}
\newcommand\Meanh[1]{\left\langle#1\right\rangle_{xy}}
\renewcommand{\vec}{\bm}
\newcommand{\kms}{\,{\rm km}\,{\rm s}^{-1}}
\newcommand{\km}{\,\text{km}}
\newcommand{\kpc}{\,{\rm kpc}}
\newcommand{\cm}{\,{\rm cm}}
\newcommand{\g}{\,{\rm g}}
\newcommand{\s}{\,{\rm s}}
\newcommand{\K}{\,{\rm K}}
\newcommand{\muG}{\,{\upmu\rm G}}
\newcommand{\yr}{\,{\rm yr}}
\newcommand{\Gyr}{\,{\rm Gyr}}
\newcommand{\betacr}{\beta_\text{cr}}
\newcommand{\betagas}{\beta_\text{m}}
\newcommand{\ecr}{\epsilon_\text{cr}}
\newcommand{\A}{_\text{A}}
\newcommand{\cra}{_\text{cr}}
\newcommand{\eff}{_\text{eff}}
\newcommand{\therm}{_\text{th}}
\newcommand{\m}{_\text{m}}
\newcommand{\mini}{_\text{min}}
\newcommand{\sound}{_\text{s}}
\newcommand{\p}{\,\text{pc}}
\newcommand{\ecri}{\epsilon_{\rm{cr0}}}

\DeclareMathOperator{\sign}{sign}
\newcommand{\rms}{{\text{r.m.s.}}}
\newcommand{\const}{{\text{const}}}
\newcommand{\Rey}{\text{Re}}

\definecolor{burntorange}{rgb}{0.8, 0.33, 0.0}

\definecolor{brown}{rgb}{0.63, 0.17, 0.0}

\definecolor{coralred}{rgb}{0.765625, 0.1171875, 0.2265625}

\definecolor{magenta}{rgb}{1.0,0.0,1.0}

\title{Steady states of the Parker instability}
\author[D.~Tharakkal et al.]{Devika Tharakkal$^{1}$, Anvar Shukurov$^{1}$\thanks{E-mail: anvar.shukurov@ncl.ac.uk}, Frederick A.~Gent$^{2,1}$,   Graeme R.~Sarson$^{1}$,
  \newauthor Andrew P.~Snodin$^{3,1}$ and Luiz Felippe S.~Rodrigues$^{4,5}$\\
$^{1}$School of Mathematics, Statistics and Physics, Newcastle University, Newcastle upon Tyne, NE1 7RU, UK\\
$^{2}$Astroinformatics, Department of Computer Science, Aalto University, PO Box 15400, FI-00076 Espoo, Finland\\ 
$^{3}$UKAEA, Culham Science Centre, Abingdon, OX14 3DB, UK\\ 
$^{4}$HAL24K Agri, Uitmeentsestraat 19, 6987 CX Giesbeek, Netherlands\\
$^{5}$Institute for Mathematics, Astrophysics and Particle Physics, Radboud University, P.O. Box 9010, 6500 GL Nijmegen, Netherlands}

\date{Accepted XXX. Received YYY; in original form ZZZ}
\pubyear{2023}

\begin{document}
\label{firstpage}
\pagerange{\pageref{firstpage}--\pageref{lastpage}}
\maketitle

\begin{abstract}
We study the linear properties, nonlinear saturation and a steady, strongly nonlinear state of the Parker instability in galaxies. We consider magnetic buoyancy and its consequences with and without cosmic rays. Cosmic rays are described using the fluid approximation with anisotropic, non-Fickian diffusion. To avoid unphysical constraints on the instability (such as boundary conditions often used to specify an unstable background state), nonideal MHD equations are solved for deviations from a background state representing an unstable magnetohydrostatic equilibrium. We consider isothermal gas and neglect rotation. The linear evolution of the instability is in broad agreement with  earlier analytical and numerical models; but we show that most of the simplifying assumptions of the earlier work do not hold, such that they provide only a qualitative rather than quantitative picture. In its nonlinear stage the instability has significantly altered the  background state from its initial state. Vertical distributions of both magnetic field and cosmic rays are much wider, the gas layer is thinner, and the energy densities of both magnetic field and cosmic rays are much reduced. The spatial structure of the nonlinear state differs from that of any linear modes. A transient gas outflow
is driven by the weakly nonlinear instability as it approaches saturation.
\end{abstract}


\section{Introduction}\label{sec:intro}

The magnetic buoyancy (or magnetic Rayleigh--Taylor) instability \citep{Newcomb61}, modified and enhanced by cosmic rays, is known as the Parker instability \citep{Parker1958,Parker1966,P79}.
The horizontal magnetic field in a gas layer confined by gravity can be
unstable with respect to undular modes which grow exponentially on a timescale
comparable to the sound or Alfv\'en crossing time over the gas density scale
height.
For the observed scale height $0.5\kpc$ for the warm interstellar gas in the 
Solar neighbourhood, with the sound and Alfv\'en speeds both at about $10\kms$,
the time scale is of order $5\times10^7\yr$.
It is much shorter than the galactic life-time, and several effects have been
explored that can make the galactic discs less unstable including the roles of
cosmic ray diffusion \citep{KP83,K87,HZ18} and rotation \citep{ZK75,FT94,FT95}.
We note in this connection that widely used heuristic models of the stratified
interstellar medium, informed by observations of the interstellar medium, are
likely to be unstable \citep[e.g.,][]{L-RCAP80}. The Parker instability can
hardly be completely suppressed in spiral galaxies, so it is important to
explore its nonlinear, steady states in order to understand why the gas
distributions observed in spiral galaxies are not destroyed by the instability. 

Aside from its effect on the vertical distributions of the interstellar gas,
magnetic field and cosmic rays, the Parker instability plays a significant role
in the evolution of galaxies. It contributes to driving galactic outflows
(winds and fountains) and thereby to what is known as the `star
formation feedback' on the evolving galactic disc.

The linear stage of the instability has been thoroughly studied and the dispersion relation has been obtained for a wide range of physical models and parameter regimes \citep[e.g.,][and references therein]{Giz1993,FT94,FT95,Kim1997,ShSu22}.

However, the nonlinear state of the Parker instability is much less
understood, in particular because it can only be studied numerically
\citep{KRJ01}. Two-dimensional (2D) simulations of the magnetic buoyancy instability
of \citet{Matsumoto1990} and \citet{1988PASJ...40..147H} show how the
instability saturate differently across different parameter ranges.
We compare our three-dimensional (3D) results with these studies, and find that 
the solutions may be rather different from any 2D linear modes and their superpositions.
 
\citet{Matsumoto1990} find two types of nonlinear 2D solutions;
oscillatory for the azimuthal (along the unstable magnetic field) wave number
$k_y$ exceeding some critical value, or shock-wave dominated for $k_y$ 
less than critical. 
When the initial magnetic pressure is much smaller than the thermal pressure
($P\m/P\therm<0.3$), the nonlinear oscillations couple to form long waves
which eventually decay due to shocks. 3D simulations described
here do not reproduce such oscillations or shocks at scales comparable to the
instability scales, and the proportions of magnetic and
cosmic ray energy density and pressure decrease over time near the midplane.
The final nonlinear stages of our models show the magnetic field
loops to have less ordered, more small-scale structure in the $xy$-plane than
these 2D simulations. \citet{Hanasz2000} and \citet{Hanasz2002} study the nonlinear
Parker instability with rotation in three dimensions, focusing on the evolution
of the magnetic field structure and the mean-field dynamo driven by the
instability, adopting a very weak initial magnetic field. \citet{Hanasz2002}
observe a decrease in the magnetic field strength at later times (following its
growth due to the dynamo action) with an increase in its scale height.
\citet{Machida2013} study the 3D nonlinear magnetic buoyancy
instability (in the absence of cosmic rays) in toroidal geometry. These authors
also observe a gradual reduction in the magnetic field strength accompanied by
an increase in its scale, possibly slowed down by the dynamo action. Both
\citet{Hanasz2002} and \citet{Machida2013} report the development of vertical
current filaments in the late stages of the instability, which could be due to a strongly nonlinear magnetic dissipation in the former case and the ideal magnetohydrodynamic approximation in the latter case. We do not observe any such features.

We explore the instability systematically, starting with the magnetic buoyancy instability and then, to clarify their role, adding cosmic rays in the fluid approximation (based on the advection--diffusion equation). Using our linear stage results where the perturbations remain weak, we test various approximations used in the linear stability analyses (which rely on widely diverse assumptions and simplifications).  In order to explore the nonlinear states of the instability, we consider \textit{imposed} and fixed background gas, magnetic field and cosmic ray distributions and solve fully nonlinear equations for the
deviations from that state.

The paper is structured as follows. In Section~\ref{sec:Basic_equations}, we set out the basic equations (Section~\ref{sec:equations}) and specify the physical and numerical parameters used in the simulations. In Section~\ref{sec:Theoretical_model}, we provide
a brief description of the gravitational field models used, and in Section \ref{Boundary_condition}, the boundary conditions are discussed. The results are presented in Sections~\ref{LInst} and \ref{NLInst} for the linear and nonlinear stages of the instability, respectively (Sections~\ref{LInst} also contains a discussion of the assumptions on which analytical studies rely). Section~\ref{discussion} contains a discussion of the implications of our results with emphasis on the nonlinear, statistically steady state, including the relative distributions of the gas, magnetic fields and cosmic rays (Section~\ref{CCBED}),  systematic vertical flows (Section~\ref{VF} and the force balance (Section~\ref{FB}). Section~\ref{sec:Conclusions} summarises our conclusions. Appendix~\ref{sec:background} justifies our approach to the implementation of the unstable background state, while the parameter space is explored in Section~\ref{StP} where we discuss the influence on the results of the scale heights of the gas and non-thermal pressure components in the background state, of the gravity profile and the role of the gas viscosity and magnetic diffusivity.

\section{Experimental design}\label{sec:Basic_equations}
As the initial state, we consider a plane-parallel magnetohydrostatic equilibrium in the galactic gravitational field, i.e., a stratified layer of thermal gas, horizontal magnetic field and cosmic rays. The instability is caused by the magnetic buoyancy, which depends on the vertical gradient of the magnetic field strength \citep{HC87}, and so leads to the removal of the initial magnetic field from the layer's midplane, tending to reduce its gradient. As a result, the initial magnetic field is rapidly lost from the system. In galaxies and accretion discs, the large-scale magnetic field is continuously replenished by the dynamo action near the disc's midplane \citep{ShSu22}. However, most simulations of the Parker instability do not include the dynamo action as the source of the unstable magnetic field. Therefore, an equilibrium state introduced as an initial condition is rapidly destroyed by the instability in such simulations. This has prevented \citet{Rodrigues2016} from analysing the steady, strongly nonlinear state of the system. Alternatively, boundary conditions can be used to impose a steady background state. However, this would constrain unphysically the evolution of the system as the fixed boundary conditions would require that the deviations from the background state vanish at the boundaries.

Therefore, our approach is to derive and solve (fully nonlinear) equations for deviations from the background state. In fact, this is the standard approach to explore the linear Parker (or any other) instability analytically, but we extend it to capture a fully nonlinear evolution of the perturbations when their magnitude is no longer small. The boundary conditions for the deviations are not restrictive (we use periodic boundary conditions in the horizontal planes), so that the perturbations can evolve freely. In the nonlinear state of the instability, the magnitude of the deviations from the background state is comparable to that of the background state, altering it fundamentally; so it is important to make the model fully flexible to allow for the possibility of such a strong modification.

In this paper, we consider an isothermal gas and neglect rotation to establish a reference model to allow us to identify the effects of radiative cooling, rotation, dynamo action and supernova activity with the associated multi-phase gas structure, which will be discussed elsewhere.

\subsection{Basic equations}\label{sec:equations}
We solve numerically the non-ideal MHD equations for the gas density $\rho$,
its velocity $\vec{U}$, total pressure $P$ (which includes the thermal,
magnetic and cosmic-ray contributions), magnetic field $\vec{B}=\nabla\times\vec{A}$, its vector potential $\vec{A}$ (with the advective gauge $\Phi=\eta\nabla\cdot\vec{A}$)  and the energy density of cosmic rays $\ecr$: 
\begin{align}
\dderiv{\rho}{t}&= - \rho \nabla \cdot \vec{U} \,,\label{cont}\\
\dderiv{\vec{U}}{t}& = -\frac{\nabla  P}{\rho} + \vec{g}+
\frac{(\nabla\times\vec{B})\times\vec{B}}{4\pi\rho} 
+\frac{\nabla\cdot\tau\,,\label{N-S}}{\rho}\\
\deriv{\vec{A}}{t}&=\vec{U}\times (\nabla\times\vec{A})
-\eta\nabla\times \nabla\times\vec{A} \label{ind}  \,,\\
\deriv{\ecr}{t}&=-\vec \nabla \cdot \left(\ecr\vec{U}\right) -P\cra \nabla\cdot \vec{U}  - \nabla \cdot \vec{F}\,,\label{ecr}
\end{align}
where $\text{D}/\text{D}t= \partial/\partial t+\vec{U}\cdot\nabla$ is the Lagrangian derivative, $\vec{g}$ is the gravitational acceleration, $P\cra=(\gamma\cra-1)\epsilon\cra$, with $\gamma\cra=4/3$ for the ultra-relativistic cosmic rays, and $\vec{F}$ is the diffusive flux of cosmic rays discussed below. The viscous term in equation~\eqref{N-S} has the form $\rho^{-1}\nabla\cdot\tau=\nu \left[ \nabla^2\vec{U}  + \tfrac{1}{3} \nabla\left(\nabla\cdot\vec{U}\right)+ 2{\mathbfss W}\cdot\nabla\ln\rho \right]$, where $\mathbfss{W}$ is the trace-less rate of strain tensor, and we use the cgs system with $c$ the speed of light. The advection--diffusion equation for cosmic rays \eqref{ecr} is derived by, e.g., \citet{Skilling1975} and \citet{Drury1981} and used in this form by, e.g., \citet{Gupta2021} and \citet{Rodrigues2016}.

Each variable is represented as the sum of a background equilibrium value, identified with the subscript zero, and a deviation from it, denoted with the prime or a lower-case symbol,
\begin{equation}\label{eq:pert}
\begin{split}
\rho&=\rho_0+\rho'\,, \quad\hspace{0.8em} \vec{U}=\vec{U}_0+\vec{u}\,, \quad\hspace{0.8em} P=P_0+P'\,, \\
\vec{B}&=\vec{B}_0+\vec{b}\,, \quad\hspace{0.8em} \vec{A}=\vec{A}_0+\vec{a}\,,\quad \hspace{0.8em}
        \ecr= \ecri+\ecr'\,,\\
\vec{F}&=\vec{F}_0+\vec{F}'\,,\quad \hspace{0.2em} \vec{U}_0=\vec{0}\,.
\end{split}
\end{equation}
The background state and how it is supported in a steady state is described in detail in Section~\ref{sec:Theoretical_model}. It is convenient to include only the deviations in the gas and cosmic ray pressures into $P'$, whereas the magnetic field deviations contribute to both the magnetic pressure and magnetic tension and thus are included as a part of the Lorentz force. However, the background pressure $P_0$ includes all three pressure components. We stress again that the deviations from the equilibrium state are not assumed to be weak: the nonlinear governing equation for them are obtained by subtracting those for the background state from equations~\eqref{cont}--\eqref{ecr} as discussed in Appendix~\ref{sec:background}, where we also present equations governing the background state which ensure that it is static, $\partial/\partial t =0$. The background magnetic field can be thought of as being maintained by the dynamo action in the disc; in this case, the assumption of its time-invariance is justified if the dynamo time scale is much shorter than the growth time of the Parker instability. The development of the Parker instability when the dynamo time scale is comparable to or longer than the instability time scale will be discussed elsewhere. 

In the absence of a background velocity $\vec{U}_0$, equations~\eqref{cont}--\eqref{ind}, written for the deviations from the background state, reduce to (Appendix~\ref{sec:background})
\begin{align}
\label{eq:continuity_pert} 
\deriv{\rho'}{t} &= - \vec{u} \cdot \vec{\nabla} \rho_0 - \rho_0 \vec{\nabla} \cdot \vec{u} - \vec{u}
\cdot \vec{\nabla} \rho^\prime - \rho^\prime \vec{\nabla} \cdot \vec{u}\,,\\
\rho \dderiv{\vec u}{t} &= \rho^\prime \vec{g}  -
\vec \nabla  P^\prime + 
\frac{(\nabla\times\vec{B})\times\vec{B}-(\nabla\times\vec{B}_0)\times\vec{B}_0}{4\pi\rho} \nonumber\\ 
&\mbox{}\quad+\rho \nu \left[ \nabla^2 \vec u  + 
    \tfrac{1}{3} \nabla \cdot \left( \nabla \cdot \vec u \right) + 
    2 {\mathbfss W} \cdot \nabla \ln \rho \right], \label{eq:mom_pert} \\
\label{eq:induction_pert}
\deriv{\vec{a}}{t} &= \vec{u} \times 
\nabla\times\vec{A} - \eta \nabla \times  \nabla\times\vec{a}\,,
\end{align}
where we recall that the pressure perturbation $P'$  contains only the gas and cosmic ray pressure perturbations (when the latter is included); the magnetic pressure perturbation is accounted for within the Lorentz force.

\label{sec:equations_cr}

Equation~\eqref{ecr} leads to the following equation for the deviation of the cosmic ray energy density from the background state:
\begin{equation}
\label{eq:crfluid_pert}
\frac{\partial \epsilon\cra'}{\partial t} =
-\vec \nabla \cdot \left(\epsilon_{\rm cr} \vec{u}\right) -P_{\rm cr} \vec
\nabla \cdot \vec{u} - \nabla \cdot \vec{F^\prime},
\end{equation}
where 
$\vec F'$ is the evolving cosmic ray flux perturbation derived from the non-Fickian diffusion model for cosmic rays \citep{Snodin+2005, Rodrigues2016} as (see Appendix~\ref{sec:background} for the derivation) 
\begin{equation}
\label{eq:crflux_pert}
    \tau \deriv{F_{i}'}{t } = -\left( \kappa_{ij} 
     \deriv{\ecr}{x_j}  
     - \kappa_{0_{ij}} 
    \deriv{\epsilon_\text{cr0}}{x_j} \right) - F_i'\,,
\end{equation}
where $\kappa_{ij}$ and $\kappa_{0_{ij}}$ are, respectively, the total and background  diffusion tensors,
\begin{align}
    \kappa_{ij} &= \kappa_{\perp} \delta_{ij} + (\kappa_{\parallel}- \kappa_{\perp})\hat{B}_{i} \hat{B}_{j}\,,\\
    \kappa_{0_{ij}} &= \kappa_{\perp} \delta_{ij} + (\kappa_{\parallel}-\kappa_{\perp})\hat{B}_{0_i} \hat{B}_{0_j},
\end{align}
where the summation convention is understood, $\hat{B}_i=B_i/|\vec{B}|$ and
$\hat{B}_{0_i}=B_{0_i}/|\vec{B}_0|$ are the components of the unit vectors in the
direction of $\vec{B}$ and $\vec{B}_0$, respectively, and the tensors
$\kappa_{ij}$ and $\kappa_{0_{ij}}$  have the same constant diffusivities
$\kappa_\parallel$ and $\kappa_\perp$, respectively parallel and perpendicular
to the relevant magnetic field. The non-Fickian description of the cosmic ray
diffusion ensures their finite propagation speed controlled by the parameter
$\tau$ (this may be important given the very large cosmic ray diffusivity), and
prevents numerical singularities in the cosmic ray propagation near the
magnetic null points \citep{Snodin+2005}. The parameters that control the
diffusion cosmic rays parameters are adopted as $\tau =0.01 \Gyr$,
$\kappa_\perp= 3.16 \times 10^{25} \, \rm cm^2s^{-1}$ and $\kappa_\parallel=
1.58 \times 10^{28} \, \rm cm^2s^{-1}$ as given in \citet[and references
therein]{Rodrigues2016,Ryu2003}. We also include a weak isotropic component of the diffusion tensor $\kappa_0=0.5\kappa_\perp$ to allow for unresolved random magnetic fields (in fact, this part of the diffusion tensor hardly affects the solutions).
Since $\vec{U_0}=\vec{0}$, equation~\eqref{eq:crfluid_pert} follows as the difference between equations~\eqref{eq:crfluid_tot} and \eqref{eq:crfluid_imp} given in Appendix~\ref{sec:background}.

\subsection{The background state}\label{sec:Theoretical_model}
The background state represents the magnetohydrostatic equilibrium in the galactic gravitational field,
\begin{equation}
\label{eq:MHE}
\nabla P_0=\rho_0 \vec{g} \,,
\end{equation}
where the total pressure consists of the thermal gas pressure $P_{\rm th0}$, magnetic pressure $P_\text{m0}$ and cosmic ray pressure $P_\text{cr0}$:
\begin{equation}\label{Ptotal}
P_0 = P_{\rm th 0} + P_\text{m0} + P_{\rm cr0} \,,
\end{equation}
and we note that the magnetic tension vanishes in this state because the background magnetic field is unidirectional, $\vec{B}_0=(0,B_0(z),0)$ in the Cartesian fame $(x,y,z)$ with $\vec{g}$ aligned with the $z$-axis.

It is convenient to introduce the pressure parameters $\beta\m$ and $\beta\cra$ as the ratios of magnetic $P\m$ and cosmic ray $P\cra$ pressures to the thermal pressure $P\therm$: 
\begin{equation}\label{eq:pratio}
 \beta\m = \frac{P\m}{P\therm} 
 \quad\text{and}\quad 
 \beta\cra = \frac{P\cra}{P\therm}\,,
\end{equation}
where 
\begin{equation}\label{eq:pressures}
P\therm = c\sound^2 \rho\,,
\quad
P\m = \frac{B^2}{8 \pi}\,,
\quad
P\cra = (\gamma\cra-1) \ecr\,.
\end{equation}
 
We consider isothermal gas. In many of our simulations (as listed in
Table~\ref{tab:my_label}), the sound speed $c\sound$ is $18\kms$ corresponding
to the temperature of $T= 3.2\times10^4\K$, which is within the range for the
warm interstellar gas, but we also consider $c\sound\in[9,29]\kms$ to explore
the effect of the gas ionization (through the analogue of sound 
speed) on the instability. In the background state, $\betagas$ and $\betacr$
are constant as listed in Table~\ref{tab:my_label}, but these ratios vary in space and time for the
deviations from the background state as the instability develops.
We mainly consider two combinations of these parameters;
$\betagas=1$ and $\betacr=0$, or $\betagas=\betacr=0.5$.

\begin{table*}
\caption{The list of simulations with the details of the gravitational field profile and parameters used, including  the magnetic field strength $B_0(0)$ and cosmic ray energy density $\ecri(0)$ in the background state at $z=0$. The pressure ratios in the background state $\betagas$ and $\betacr$ are defined in equation~\eqref{eq:pratio} and $h$ is the effective exponential height of the gas in the background state. The viscosity $\nu$ and magnetic diffusivity $\eta$ are given as the multiples of $1\kpc\kms=3\times10^{26}\,\rm cm^2\,s^{-1}$, and $c\sound$ is the sound speed. The grid spacing along the $x$, $y$ and $z$-axes are given in the penultimate column, with the computational box size of $(4\times4\times3.5)\kpc^3$.}
     \centering
\begin{tabular}{cccccccccc}
\hline
&$g(z)$ &$h$    &$\betagas$ &$\betacr$  &$B_0(0)$   &$\nu$ &$\eta$  &$c\sound$  &$(\Delta x, \Delta y, \Delta z)$\\
&       &[kpc]  &           &           &[$\upmu$G] &\multicolumn{2}{c}{[kpc\,km\,s$^{-1}$]}    &[km\,s$^{-1}$] &[pc]\\
\hline
Sim1 &Eq.~\eqref{gtanh}     &0.5    &1.0    &0      &7  &0.1    &0.03   &18 &(15, 7, 13)\\
Sim2 &Eq.~\eqref{gtanh}     &0.5    &0.5    &0.5    &5  &0.1    &0.03   &18 &(15, 7, 13)\\
Sim3 &Eq.~\eqref{gtanh}     &1.0    &0.5    &0.5    &7  &0.1    &0.03   &25 &(31, 15, 27)\\
Sim4 &Eq.~\eqref{gtanh}     &1.0    &0.25   &0.25   &4  &0.1    &0.03   &29 &(31,15, 27)\\
Sim5 &Eq.~\eqref{KG89grav}  &0.4    &1.0    &0      &7  &0.1    &0.03   &18 &(15, 7, 13)\\
Sim6 &Eq.~\eqref{KG89grav}  &0.4    &0.5    &0.5    &5  &0.1    &0.03   &18 &(15, 7, 13)\\
Sim7 &Eq.~\eqref{KG89grav}  &0.4    &0.5    &0.5    &5  &1.0    &0.3    &18 &(31, 15, 27)\\
Sim8 &Eq.~\eqref{KG89grav}  &0.3    &1.0    &0      &3  &0.1    &0.03   &9  &(31, 15, 27)\\
Sim9 &Eq.~\eqref{KG89grav}  &0.4    &0.5    &0.5    &5  &0.1    &0.03   &18 &(7, 7, 7)\\
\hline
\end{tabular}
\label{tab:my_label}
\end{table*}

For a given gravitational profile $g(z)$ we solve equation \eqref{eq:MHE} to yield the
corresponding density, magnetic field and cosmic ray profiles, given the constants $\betagas$ and $\betacr$. 
Following some studies of the Parker instability \citep[e.g.,][]{Giz1993,Rodrigues2016} we use the gravitational acceleration profile of the form 
\begin{equation}\label{gtanh}
g(z)=-2\pi G\Sigma \tanh{(z/H)}\,,
\end{equation}
with  $H=500\,\rm pc$, $\Sigma=10^2\,\rm M_\odot\p^{-2}$ for the surface mass density and $G$ Newton's gravitational constant, which applies to the self-gravitating stellar disc. This gravity profile is used for the sake of comparison with the earlier results but most of our results are based on the gravity field appropriate for the Solar vicinity of the Milky Way, which also includes the contribution from the dark matter halo \citep{1989MNRAS.239..571K},
\begin{equation}\label{KG89grav} 
g(z) = -\frac{a_1z}{\sqrt{z_1^2+z^2}}-\frac{a_2z}{z_2}\,, 
\end{equation} 
where $a_1 = 4.4 \times 10^{-9}\, \rm  cm\,s^{-2}$, $a_2 = 1.7 \times 10^{-9}\, \rm  cm\,s^{-2}$, $z_1=0.2 \, \rm kpc$ and $z_2 = 1 \, \rm kpc$. \citet{LBS17} propose an alternative form for the gravitational acceleration better applicable at $|z|\gtrsim2\kpc$, while \citet[][equation 35]{1998ApJ...497..759F} suggests an extension of equation~\eqref{KG89grav} to other galactocentric distances. For the gravity profile \eqref{KG89grav}, the isothermal gas density, the strength of the magnetic field (directed along the $y$-axis) and cosmic ray distribution in the magnetohydrostatic equilibrium follow as (see Appendix~\ref{sec:background})
\begin{align} 
\rho_0(z) &= \rho_0(0) \exp\left(\!a_1\frac{z_1-\sqrt{z^2+z_1^2} 
- a_2z^2/(2a_1z_2)}{ c_s^2 (1+\beta\m +\betacr)}\!\right)\!\!,\\
\label{B0z}
B_0(z)&=\deriv{A_{0x}}{z}= [8\pi\beta\m c\sound^2\,\rho_0(z)]^{1/2}\,,\\
\label{ecr0z}
\epsilon_\text{cr0}(z)&=\tfrac{1}{3} \betacr c\sound^2 \rho_0(z)\,,
\end{align}
where we adopt $\rho_0(0) =7 \times 10^{-25} \, \rm g\, cm^{-3}$. Although the density distribution in $z$ deviates from an exponential, it is useful to characterise it using the fitted exponential scale height $h$ given in Table~\ref{tab:my_label}, where we summarise the model parameters used in the simulations.

Similar expressions for the gravity profile \eqref{gtanh} can be found in \citet{Rodrigues2016}. Although the qualitative picture of the development and saturation of the instability is the same under both gravity profiles \eqref{gtanh} and \eqref{KG89grav}, there are important quantitative differences in both the linear eigenmodes and some features of the nonlinear state: in particular, the outflow speed.

\begin{figure}
    \centering
    \includegraphics[width=0.85\columnwidth]{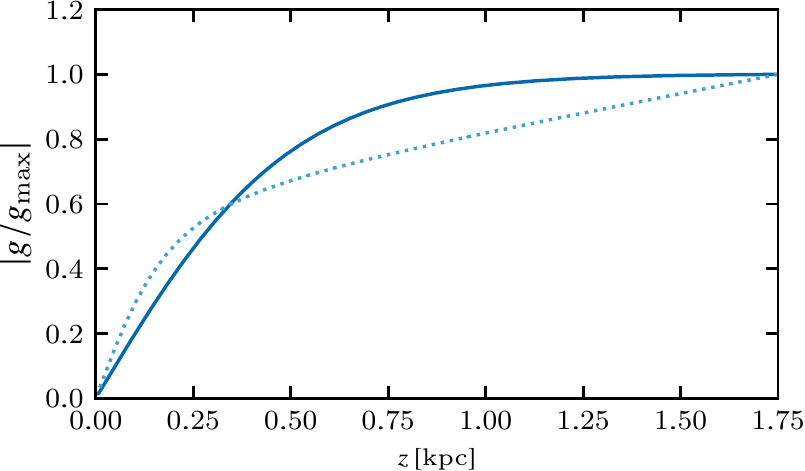}
    
    \caption{The magnitudes of the gravitational acceleration, Eq.~\eqref{gtanh} (solid) and Eq.~\eqref{KG89grav} (dotted), normalized to their magnitude at $z=Z_0$.
    }
    \label{fig:gravprofs}
\end{figure}

The gravity profile \eqref{gtanh} has been adjusted to provide similar values
of $g_z$ to those of \eqref{KG89grav} within the computational domain used
here, $|z|\leq Z_0=1.75\kpc$ (Fig.~\ref{fig:gravprofs}), so we do not expect any
strong differences in the solutions obtained under the two profiles. However,
some rather subtle differences may still occur.

The background magnetohydrostatic equilibrium is unstable. A weak initial velocity perturbation is introduced to launch the instability and the perturbations are then evolved using equations \eqref{eq:continuity_pert}--\eqref{eq:crfluid_pert}.

 \subsection{Boundary conditions and numerical implementation} \label{Boundary_condition} 

We use the \textsc{Pencil Code} \citep{Pencil-JOSS} to simulate a local
Cartesian box within the gas layer; here we neglect rotation \citep[for effects of rotation see][]{TSGSS22}. We simulate magnetised interstellar gas and cosmic rays
in a region of the size $4\times4\times3.5\text{ kpc}^3$ (symmetric about the
midplane located at $z=0$) along the $x$, $y$ and $z$-axes, respectively, and
use periodic boundary conditions in $x$ and $y$. The extent of the domain along
the background magnetic field ($y$) is chosen to be large enough to accommodate
the dominant modes of the instability (see Section~\ref{LInst}).

The boundary conditions at the top and bottom of the domain, $z=\pm Z_0$ with $Z_0=1.75\kpc$, are given in equations~\eqref{bcA}--\eqref{bcecr} and chosen to minimise their effect on the interior of the domain while remaining physically justifiable. Their effect on the processes within the domain is weak or negligible also because $Z_0$ exceeds by a factor 2--3 the scales of the background gas, magnetic field and cosmic ray distributions.

The magnetic field perturbations satisfy the boundary conditions
\begin{equation}\label{bcA} 
\deriv{a_x}{z} = \deriv{a_y}{z}=0\,,\quad  a_z=0\,. 
\end{equation} 
This allows $B_z \neq 0$ at $|z|=Z_0$, and the open magnetic lines can support the gas and cosmic ray flows across the boundary.

The boundary condition for $\rho^\prime$ at $|z|=Z_0$ imposes an exponential 
decrease at $|z|>Z_0$ (i.e., within the ghost zones used to impose the boundary condition), at the scale that corresponds to the evolving vertical thermal pressure gradient at $|z|=Z_0$.

The horizontal velocity perturbations vanish at the top and bottom of the domain,
\begin{equation}\label{bcuxy} u_x = u_y=0 \quad\text{at } |z|=Z_0\,.
\end{equation} 
The vertical velocity at $|z|=Z_0$ should be handled carefully to allow
unrestricted gas outflow, while constraining the gas inflow to within
numerically stable limits. To suppress numerical instabilities (associated with
spurious strong advection of gas from the outside of the domain), any negative
$u_z$ at $z=Z_0$ and positive $u_z$ at $z=-Z_0$ (an inflow, $[u_z(x,y,z)\sign
z]_{|z|=Z_0}<0$) are gradually quenched to zero within the three ghost points at
$|z|>Z_0$. At those horizontal positions on the top and bottom boundaries where
the gas flows out of the domain ($[u_z(x,y,z)\sign z]_{|z|=Z_0}>0$), the
vertical derivatives
of $u_z$ of all relevant orders are assumed to vanish in the ghost zone. 
Thus, the gas that flows out through $|z|=Z_0$ retains its speed in the ghost zone.
These conditions, imposed within the ghost zone rather than on the boundary, can be written as
\begin{equation}\label{bcuz}
\left.
\begin{split}
\text{inflow:\quad}& u_z\to0 \\
\text{outflow:\quad}& \deriv{^n u_z}{z^n}=0 \text{ for any $n$}
\end{split}
\right\}\quad \text{at } |z|>Z_0\,,
\end{equation}
where the higher-order derivatives are involved when hyperdiffusion is used.

The top and bottom boundaries are open for the cosmic ray fluid, allowing it to escape along open magnetic lines at $|z|=\pm Z_0$,
\begin{equation}\label{bcecr}
    \deriv{^2\ecr}{z^2} = 0\,,\quad 
    \deriv{F_x}{z} = \deriv{F_y}{z} =0\,,\quad F_z = 0\,.
\end{equation}

The viscosity and magnetic diffusivity are global constants in our simulations, and we considered a selection of their values shown in Table~\ref{tab:my_label}; in all cases, the magnetic Prandtl number is $P_m=\nu/\eta\approx 3$, as in \citet{Rodrigues2016}.
The magnitudes of the transport coefficients are close to (but, in all models except for Sim7, smaller than) their turbulent values in the ISM; this is appropriate since our simulations do not not include the turbulence driven by supernovae.
As the instability saturates, small-scale fluctuations get stronger making the system more susceptible to numerical instabilities.
The fluctuations at the resolution scale are regularised using  sixth-order hyperdiffusion with the diffusivity $\nu_6=4 \times 10^{-11}\, \rm kpc^6\, Gyr^{-1}$ \citep[e.g.,][]{ABGS02,HB04}, constraining the maximum mesh Reynolds number to $\Rey_\Delta=c\sound\Delta x^5/\nu_6$. The magnitude of $\nu_6$ was chosen to 
ensure that $\Rey_\Delta$ is close to
unity along the lowest-resolution dimension, $x$. Shocks arising in the system are regularised
using the second-order shock diffusion  \citep[for implementation, see][]{2020GApFD.114...77G}. To avoid negative gas density values that may arise in intense divergent flows, we apply a density floor and impose the restriction $\rho_0+\rho^\prime\geq \rho_{\rm min}$ with $\rho_{\rm min}=0.01\, \rm M_{\odot} \kpc^{-3}$. The magnitude of $\rho_\text{min}$ chosen is about one order of magnitude lower than the initial minimum density in the simulation box.

Table~\ref{tab:my_label} presents a summary of the simulations which are based
on the two forms of the gravity acceleration, given in equations~\eqref{gtanh}
and \eqref{KG89grav}, and a selection of the pressure ratios of the magnetic
field, $\betagas$, and cosmic rays, $\betacr$, to the thermal pressure in the
background state. These factors control the intensity of the instability and
the contribution of cosmic rays to it; the case $\betacr=0$ represents the
magnetic buoyancy instability.  
We list the sound speed $c\sound$ in each model.
We compare a simulation with 
the fiducial values (15,~7,~13)\p\ for the grid
spacing ($\Delta x$, $\Delta y$, $\Delta z$) along the corresponding axes,
 to a simulation with (7,~7,~7)\p.

\section{Results}\label{sec:Results}
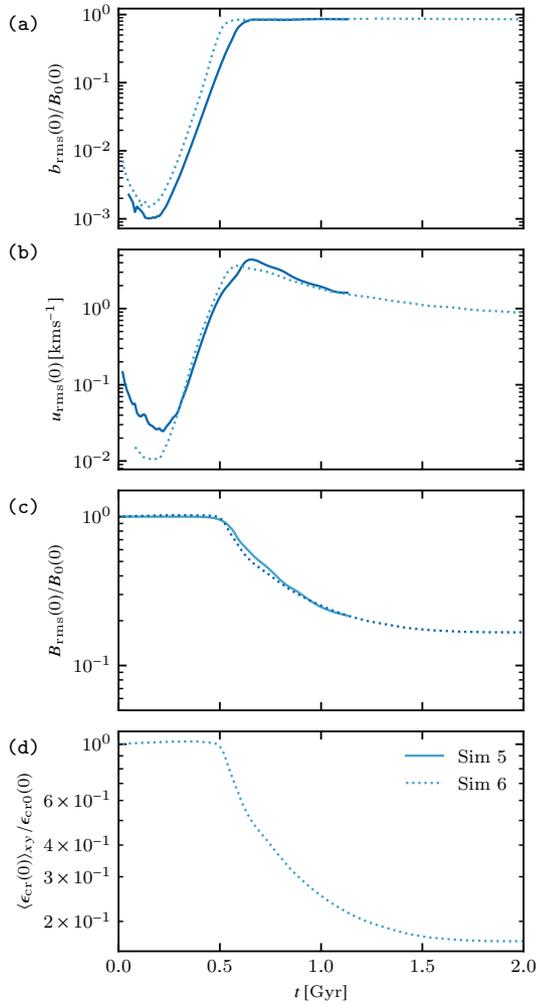
\begin{figure}
\centering 
\includegraphics[trim=0cm 0.2cm 0cm 0.2cm, clip=True, width=0.85\columnwidth]{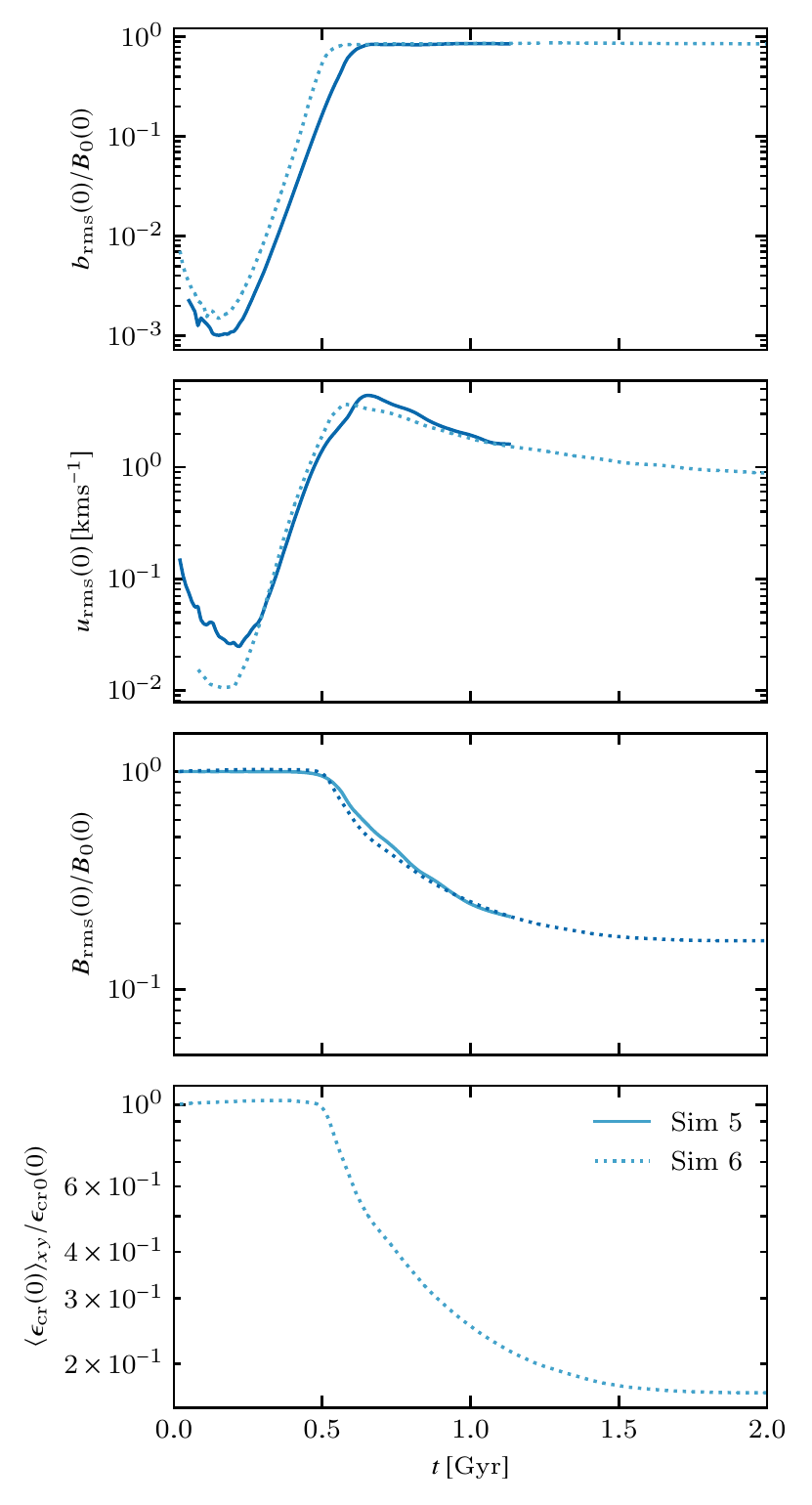}
 \begin{picture}(1,0)
 \put(0.085,1.535){\texttt{(a)}}
 \put(0.085,1.185){\texttt{(b)}}
 \put(0.085,0.795){\texttt{(c)}}
 \put(0.085,0.425){\texttt{(d)}}
 \end{picture}
\caption{The evolution of the root-mean-square magnitudes
at the midplane $z=0$ of \textbf{(a)}~the magnetic field perturbation $|\vec{b}|$, normalised to $B_0(0)$, and \textbf{(b)}~gas speed in the models Sim5 (solid) and Sim6 (dotted) at $z=0$. The linear stage of the instability ends at about $t\simeq0.4\,\rm Gyr$. Panels~\textbf{(c)} and \textbf{(d)} show the normalised total magnetic and cosmic ray energy densities at the midplane for the same models, 
 $\langle B(0)/B_0(0)\rangle_{xy}$ (solid for Sim5 and dotted for Sim6) and $\langle\ecr(0)/\ecri(0) \rangle_{xy}$ (dotted, for Sim6) respectively.
} \label{fig:kg89rmGR}
\end{figure}

\begin{figure}
\centering
\includegraphics{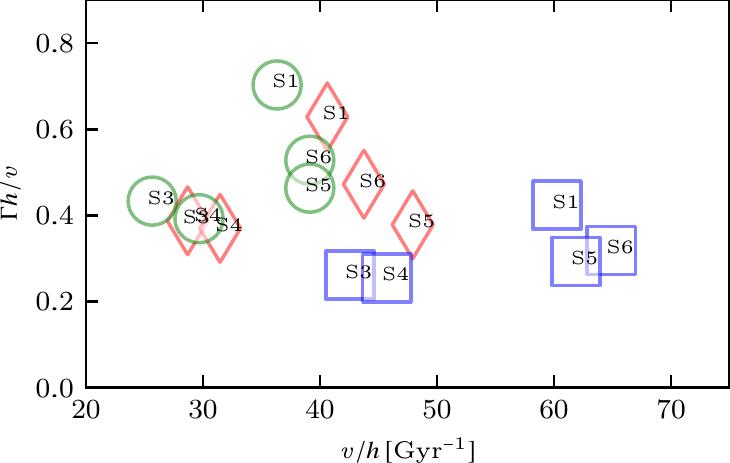}
\caption{The dimensionless growth rate $\Gamma h/v$ of the \rms\ magnetic field
	perturbation is shown for various models,
	normalized with various crossing times over the density
	scale height $h$: $v=c\sound$ (diamonds), $v=V\A$ (circles) and
	$v=c_\text{eff}$ of equation~\eqref{ceff} (squares). The horizontal axis
	presents the corresponding inverse time scale $v/h$. The model codes
	are shown within the symbols as in \autoref{tab:my_label}, e.g. S5
	represents Model Sim5.
    }
    \label{fig:gr_comp}
\end{figure}

Our main conclusions are drawn from the numerical models Sim5 and Sim6 in Table~\ref{tab:my_label}. They differ in the contribution of cosmic rays described using equations~\eqref{eq:crfluid_pert} and \eqref{eq:crflux_pert}: Sim5 is the case of the magnetic buoyancy instability, enhanced by cosmic rays in Sim6.
The total pressure in the background state remains unchanged since $\betagas+\betacr=1$ in each case. 

\subsection{The linear instability}\label{LInst}

The exponential growth in the perturbations of magnetic field and gas speed in the linear phase of the instability at $t\lesssim0.4\Gyr$ (and after the decay of the initial perturbations down to the leading eigenfunction) is clearly visible in Fig.~\ref{fig:kg89rmGR}a,b.  As shown in Fig.~\ref{fig:kg89rmGR}c, the total magnetic and cosmic ray energy densities vary little with time as long as the perturbations remain much weaker than the background fields; their subsequent nonlinear evolution is discussed in Section~\ref{NLInst}. The growth rate of the instability in the models of \autoref{tab:my_label} is quantified in Fig.~\ref{fig:gr_comp}, which presents both the dimensionless growth rate $\Gamma h/v$ and its unit $v/h$ for various choices of the characteristic speed $v$.  

In agreement with earlier analytical and numerical models, cosmic rays make the system more unstable and the growth rate is smaller in Model~Sim5 which has no cosmic rays, than in Sim6 where the cosmic rays and magnetic field have equal pressures in the background state. The growth rate $\Gamma$ of the root-mean-square (\rms) velocity and magnetic fields in these models increases by approximately $20\%$ due to the cosmic rays from $19\Gyr^{-1}$ to $25\Gyr^{-1}$, similar to the increase in $\Gamma$ in the analytical models of \citet{Giz1993} and \citet{Ryu2003} and the simulations of \citet{Rodrigues2016} (the latter authors provide a detailed comparison of the growth rates in several models).

The growth rate of the Parker instability is often assumed to scale with (and,
often, to be of the order of) the inverse sound or Alfv\'en crossing time over
the density scale height, $c\sound/h$ or $V\A/h$, respectively. Our results
show that this scaling is rather inaccurate as the growth rate depends
significantly on other parameters of the system. Figure~\ref{fig:gr_comp} shows
the dimensionless growth rates for models from Table~\ref{tab:my_label} normalised by
various crossing times, $\Gamma h/v$, where $v=c\sound$, $v=V\A$ and
$v=c\eff$ with
\begin{equation}\label{ceff}
c\eff^2= c\sound^2 + V\A^2 + \betacr c\sound^2\,.
\end{equation}
We note that the horizontal spread of the data points indicates a variation of
$\Gamma h/v$ by a factor of two between various models for any choice of $v$.
Secondly, if the scaling of $\Gamma$ with $v/h$ were perfect, the data points
would be form a horizontal line for the most appropriate choice of
$v$. This does not occur for any choice of $v$, but the scaling with $c\eff$
(shown with squares) is marginally better, especially if the Model Sim1 is
disregarded. The models that are not included in Fig.~\ref{fig:gr_comp} have similar growth rates and do not provide additional information.

\begin{figure}
\centering
\includegraphics[trim=0cm 0.25cm 0cm 0cm, clip=True, width=0.85\columnwidth]{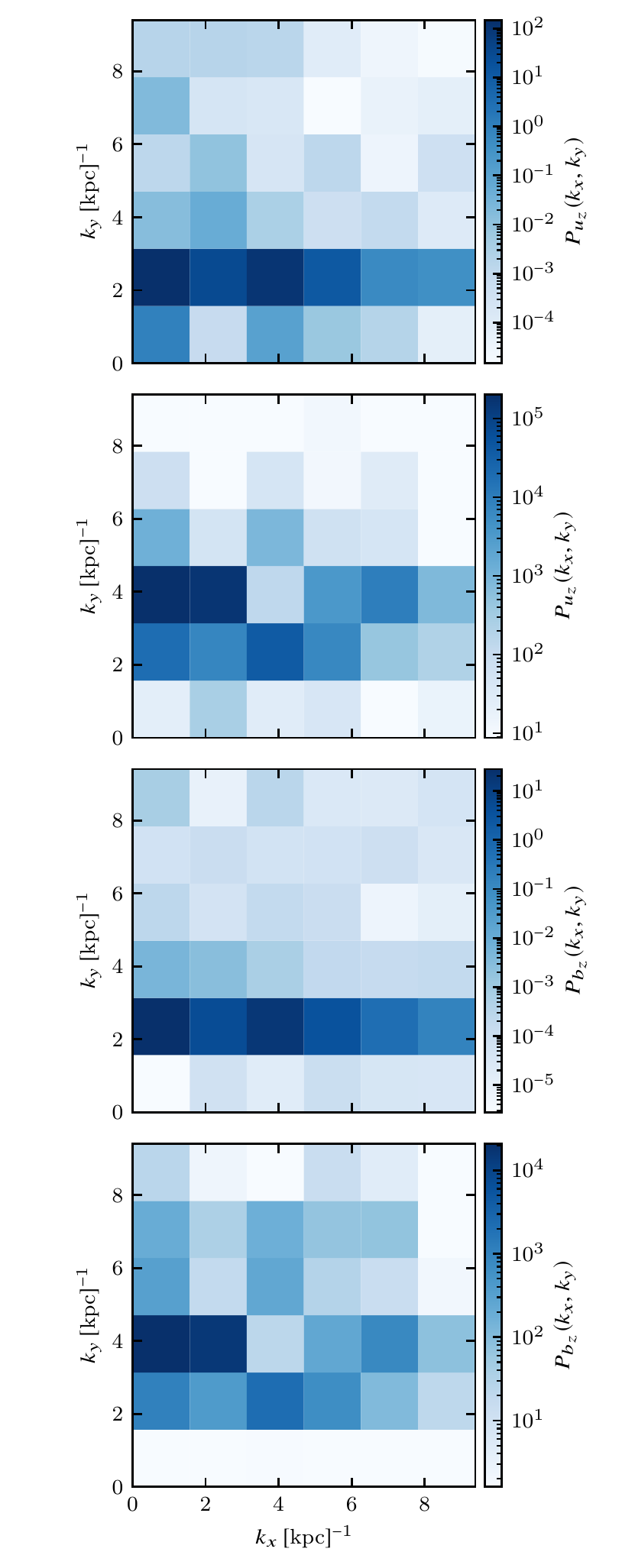}
 \begin{picture}(1.0,0)
 \put(0.16,2.070){\sf{\bf{(a)}}}
 \put(0.16,1.555){\sf{\bf{(b)}}}
 \put(0.16,1.065){\sf{\bf{(c)}}}
 \put(0.16,0.540){\sf{\bf{(d)}}}
 \end{picture}
\caption{2D power spectra in the $(k_x,k_y)$-plane at $t=0.3\Gyr$ (the
	linear stage of the instability) at $|z|\leq 1.75\kpc$ for the perturbations
	as indicated in the colour bar labels of
	vertical velocity $u_z$ in kpc$^2$km$^{2}$\,s$^{-2}$ and the
	vertical magnetic field $b_z$ in kpc$^2\muG^{2}$.
	Panels {\textbf{(a)}} and {\textbf{(c)}} depict 
	Model Sim5 and {\textbf{(b)}} and {\textbf{(d)}} present Model Sim6.}
\label{fig:2psdkg89}
\end{figure}
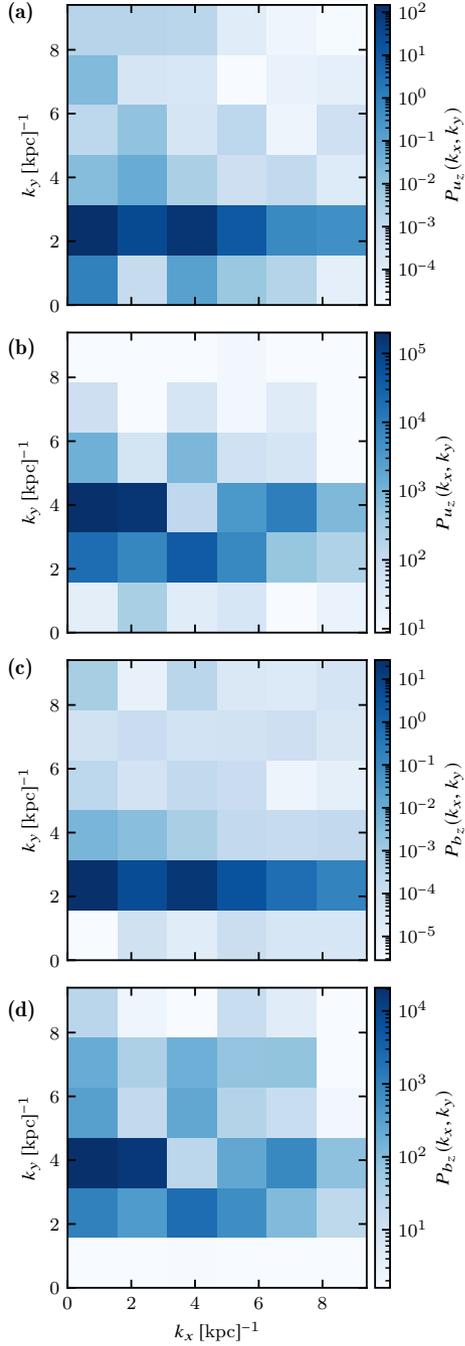 

Analytical and numerical studies of the Parker instability rely on a wide diversity of assumptions and, perhaps not surprisingly, different models claim different properties of the most rapidly growing mode (we note that many studies are based on ideal magnetohydrodynamics). For example, \citet{Parker1969} assume that the perturbations are only weakly dependent on $z$ ($k_z\to0$). The most rapidly growing mode of \citet{L-RCAP80} has $k_x=0$, $(k_y h)^2=0.5\text{--}0.6$.  \citet{Giz1993} assume that the perturbations have finite $k_x$ and $k_y$ and the amplitudes of both $b_z$ and $u_z$ vary similarly with $z$ while $|u_z|\propto \rho_0^{-1/2}(z)$ at large $|z|$; the most rapidly growing mode has a very large $k_x$. \citet{FT94} adopt $k_x\neq0$ and $k_y\neq0$ and the magnitude of the perturbations is assumed to be proportional to $\rho_0^{-1/2}(z)$. In the model of \citet{Parker1969}, the system is the most unstable at the largest possible $k_x$, and \citet{Giz1993} assume that the most unstable mode has a large radial wave number, $k_x h=10^3$ (their Fig.~2), whereas \citet{HZ18} assume that $k_x\to0$. \citet{KiRy01} argue that $k_x=k_z=0$ in the most rapidly growing mode. \citet[][Section~2.8.2]{ShSu22} assume that the perturbation amplitudes of the density, thermal pressure and Lagrangian displacement scale with $\rho_0^{-1/2}(z)$ while the amplitude of $\vec{b}$ is independent of $z$, and that the perturbations have $k_x=0$ but finite $k_y$. The wave numbers of the unstable modes along the background magnetic field, $k_y$, are limited from above \citep{FT94},
\begin{equation}\label{kP}
  k_y^2\leq   k_\text{P}^2 = (1+\beta\m+\betacr) \frac{\beta\m+\betacr}{2\beta\m}h^{-2}.
\end{equation}
For a finite magnetic diffusivity $\eta$, the radial wave number must be smaller than \citep{FT94}
\begin{equation}\label{kxmax}
    \bar{k}_x \simeq \sqrt{\frac{c\sound}{h \eta}}\,.
\end{equation}

Our simulations are consistent with some of these assumptions and conclusions. The spatial structure of the most unstable linear mode is characterised by Fig.~\ref{fig:2psdkg89} where we present the 2D power spectra of $u_z$ and $b_z$ in the $(k_x, k_y)$-plane, averaged over $|z|<1.75\kpc$, for the Models Sim5 (without cosmic rays) and Sim6 (with cosmic rays). The spectra of these variables are nearly identical. The presence of cosmic rays does not change $k_x$ much but increases $k_y$ significantly (by a factor of two). The energy corresponding to the most unstable mode is also higher by a factor of 4 in Sim6 when compared to Sim5. This confirms that cosmic rays make the system more unstable. The values of $k_x$ are well within the upper limit $\bar{k}_x\approx 40\kpc^{-1}$ of equation~\eqref{kxmax}.

In Model Sim6, $k_{\rm P} = 3.5\kpc^{-1}$, corresponding to the wavelength $\lambda_\text{P}=1.8\kpc$ (and $k_\text{P}=2.5\kpc^{-1}$ in Model Sim5).

The size of the computational domain along the $y$-axis, $4\kpc$, is larger than that, so it can accommodate the most unstable modes (see also Section~\ref{SCD} where we also discuss the wave numbers of the most unstable mode). The spectrum  of the perturbations in this model, shown in Fig.~\ref{fig:2psdkg89}b, has the maximum at $k_y\simeq 3.1\kpc^{-1}$, in agreement with equation~\eqref{kP}. 
As discussed in Section~\ref{SCD}, the power spectrum of the perturbations has a pronounced maximum at $k_y\approx3\kpc^{-1}$ but a broad range of modes at $k_x\la3\kpc^{-1}$ carry similar energies for the parameter values of Model Sim6.

\begin{figure}
    \centering
    \includegraphics[width=0.85\columnwidth]{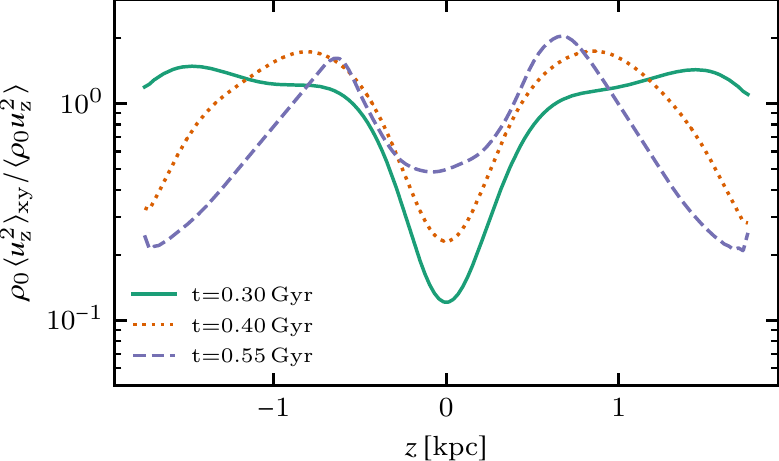}
    \caption{For Sim6 during
	the linear and weakly nonlinear stages of the instability at times indicated
	in the legend the vertical variation of $\rho_0\meanh{u_z^2}$ normalized by
	its volume averaged energy density $\langle\rho_0u_z^2\rangle$.
	}
    \label{fig:rhouz2}
\end{figure}
\begin{figure}
    \centering
     \includegraphics[width=0.85\columnwidth]{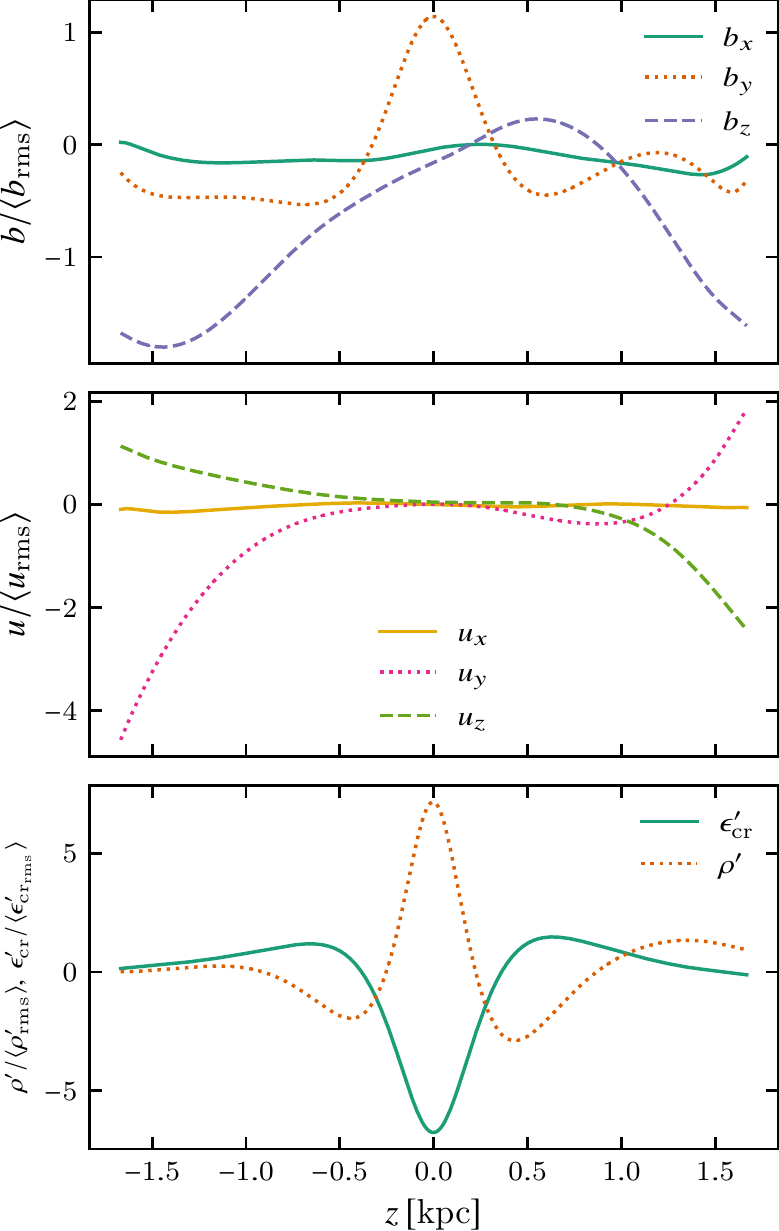}
    \begin{picture}(1.0,0)
     \put(0.07,1.350){\sf{\bf{(a)}}}
     \put(0.07,0.925){\sf{\bf{(b)}}}
     \put(0.07,0.500){\sf{\bf{(c)}}}
     \end{picture}
     \caption{The vertical profiles of the perturbation magnitudes normalised by their volume averaged r.m.s.\ values throughout the domain in Model Sim6 for different variables as denoted in the legends at $(x,y)=(-1.2,-1.8)\p$ and $t=0.3\Gyr$. The profiles at fixed $(x,y)$ are shown because the horizontal averages of the perturbations are negligible during the linear phase of the instability.
     }
    \label{fig:zvars}
\end{figure}
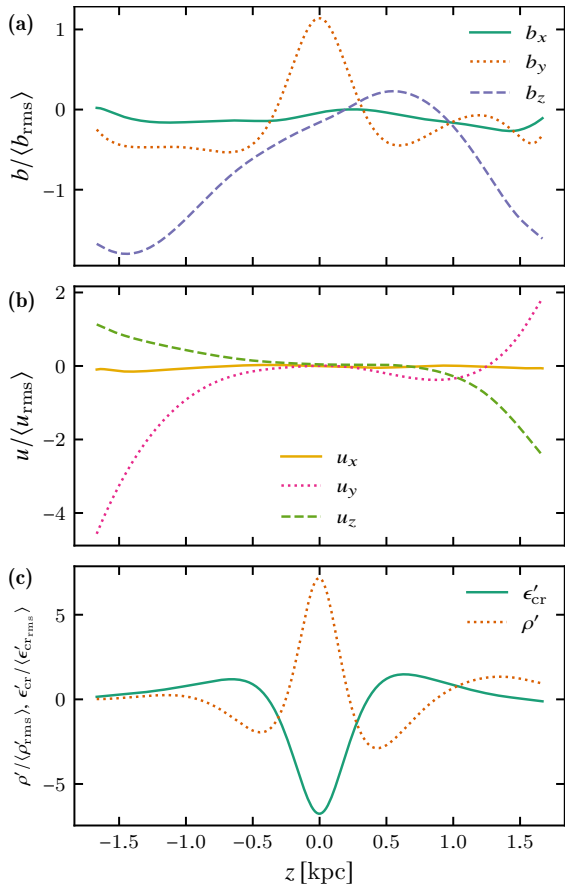

\citet{Giz1993} discuss continuous modes assuming that $\rho_0 u_z^2=\const$. Figure~\ref{fig:rhouz2} shows the variation with $z$ of the horizontal average $\rho_0\meanh{u_z^2}$ (normalized to its mean value across the computational domain; we note that $\meanh{u_z^2}$ grows exponentially with time).
\footnote{The averages at $z=\const$ (horizontal averages) are denoted $\meanh{\cdots}$ while $\langle\cdots\rangle_{xyz}$ denotes a volume average over the whole computational domain.}
The variation is by  an order of magnitude across the full range of $z$ and is not monotonic. \citet{ShSu22} consider solutions where $b$ is independent of $z$ and the perturbations in the other variables decrease as $\exp[-z/(2h)]$ (in particular, $\rho_0 u_z^2=\const$). While the perturbations in our simulations do not satisfy these assumptions closely, the one-dimensional perturbation power spectra in $z$ show that the  most unstable modes of $u_z$ and $b_z$ have $k_z \approx 3.5\kpc^{-1}$ in Models Sim5, Sim 6 and Sim9, corresponding to the scale $2\pi/k_z=1.8\kpc$. As shown in Fig.~\ref{fig:zvars}, the magnitude of perturbations in different quantities vary differently with $z$, in contradiction to the assumption of the standard stability analysis. In particular, $\rho$ and $\ecr$ have oscillations in $z$ superimposed on a large-scale variation. The perturbations in the velocity and magnetic field components $u_x$ and $b_x$ (which have relatively small amplitudes) and $u_z$ and $b_z$ (having large amplitudes) are oscillatory in $z$. Meanwhile, $u_y$ and $b_y$ do not have pronounced oscillations in $z$. The values of $k_z$ presented characterise mainly the  oscillatory parts of the variation  and should be interpreted with caution.

\begin{figure}
     \centering
    \includegraphics[trim=0.25cm 0.25cm 0.25cm 0.25cm, clip=True, width=0.95\columnwidth]{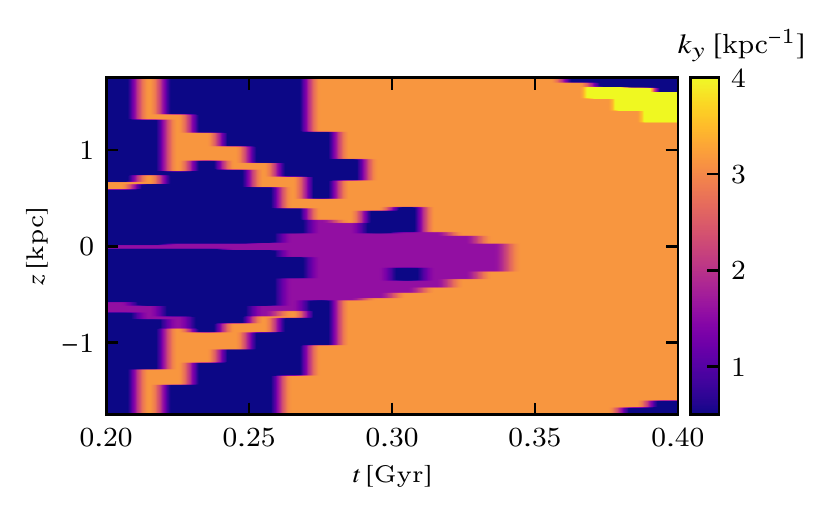}
    \caption{Time-latitude diagram of the azimuthal wave number $k_y$ for $_y$ during
	the linear instability of Model Sim6.    }
    \label{fig:kyzt}
\end{figure}


\begin{figure}
    \centering
    \includegraphics[width=0.85\columnwidth]{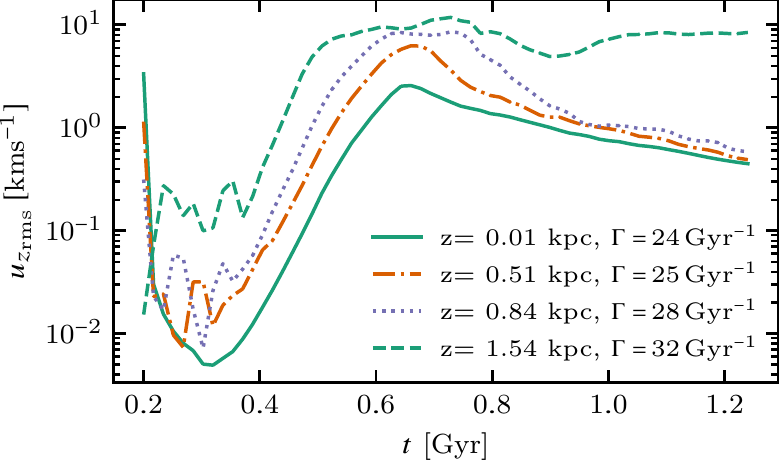}
    \caption{Model Sim6 evolution of $\meanh{u_{z_\text{rms}}}$ at altitudes
	indicated in the legend. Corresponding growth rate $\Gamma$ of the
	instability listed in the legend applies for $0.4\lesssim t \lesssim
	0.5$ Gyr. Evolution is similar for $z<0$. }
    \label{fig:uz_rms_tanh}
\end{figure}
\begin{figure*}
\includegraphics[width=1.85\columnwidth]{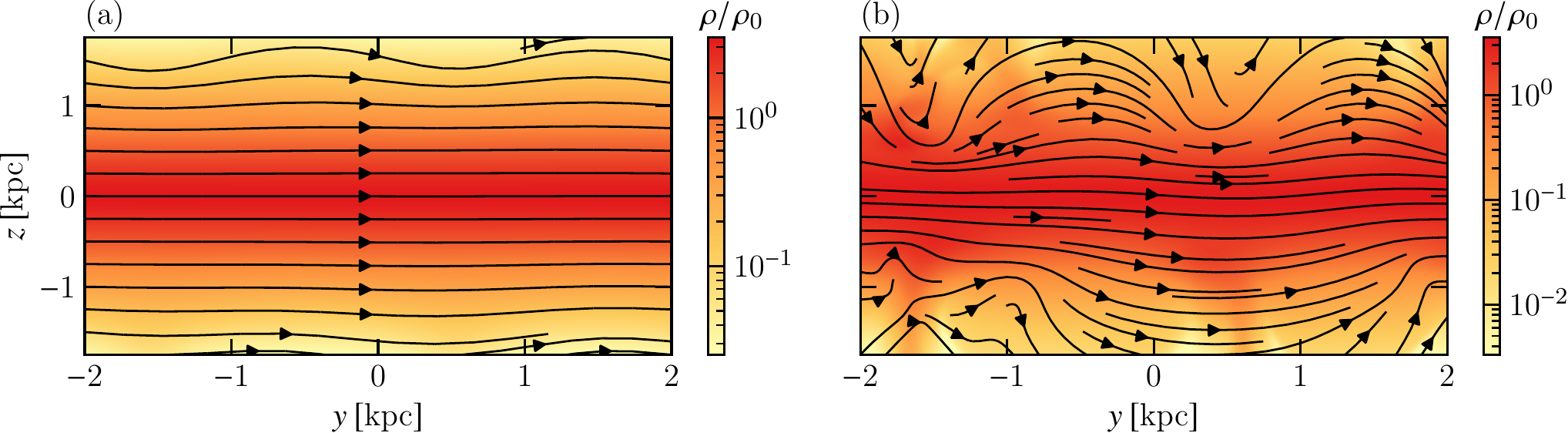}\\
\includegraphics[width=1.85\columnwidth]{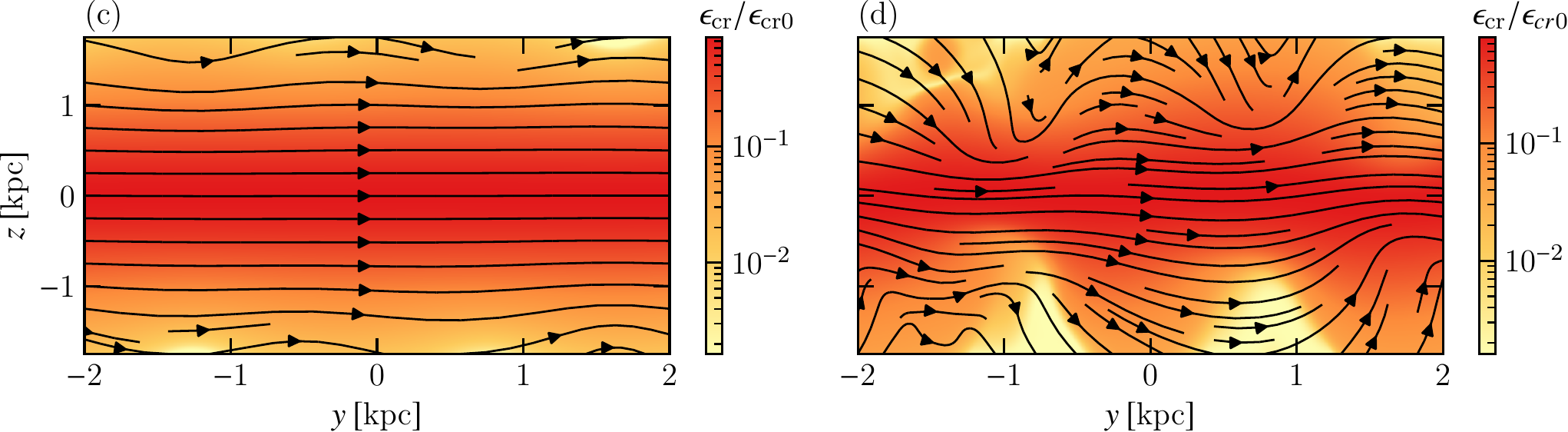}
\caption{\textbf{Upper row:} the gas density distribution (colour) and magnetic lines, with the local field direction indicated, in the cross-section at $x=800\p$ in Model Sim6  at \textbf{(a)}~$t=0.30\Gyr$ and \textbf{(b)}~$t=0.45\Gyr$. \textbf{Lower row:} as above but for the cosmic ray energy density (colour) at \textbf{(c)}~$t=0.30\Gyr$ and \textbf{(d)}~$t=0.45\Gyr$. The separation of the magnetic lines does not represent the field strength.
}
\label{fig:kg89MFNL} 
 \end{figure*}

The stratification of the system affects the spatial structure of the linearly unstable modes; in particular, their wave numbers vary with position and time. For example, Fig.~\ref{fig:kyzt} shows that the wave number $k_y$ of the dominant mode changes with both $z$ and $t$, generally increasing with time at a fixed $z$. During 
the late stages of
the exponential growth phase, $t\simeq0.3\Gyr$, $k_y$ increases from $0$ to $3\kpc^{-1}$ with distance from the midplane and approaches $k_y\approx3\kpc^{-1}$ in the nonlinear phase at all distances (see also Fig.~\ref{fig:kg89MFNL}). Correspondingly, the growth rate of the instability is no longer a constant but becomes a function of $z$. As shown in Fig.~\ref{fig:uz_rms_tanh}, the growth rate of the vertical velocity increases with $|z|$ from  $24\Gyr^{-1}$ at $z=0$ to $32\Gyr^{-1}$ at $z = 1.5\kpc$. This makes the linear evolution more complicated than the linear analytical models suggests but the properties of the dominant mode are broadly consistent with certain caveats discussed here. 

Figure \ref{fig:kg89MFNL} illustrates the development of the spatial structures of the magnetic field, gas density and cosmic rays in the linear
stage of the instability in panels (a) and (c), where the perturbations are weak at all $|z|$, and at the end of the linear stage in panels (b) and (d), where the perturbations are quite strong at $|z|\ga h=0.4\kpc$ but still are weaker closer to the midplane. The magnetic field lines are aligned with the $y$-axis initially, and the instability excites the undular modes which deform the magnetic  field lines and cause the gas and cosmic rays to be redistributed in $z$. As the system evolves into the nonlinear state, magnetic fluctuations develop at progressively smaller scales. 

Model Sim9 has a somewhat higher spatial resolution than Model Sim6 (with all other parameters being similar). The wave numbers and growth rates obtained from the two models are quite similar confirming that model Sim6 resolves well the spatial structures developing in the system. Model Sim8 is similar to Model Sim5 but has a lower sound speed (thus, a lower thermal pressure) and, consequently, a lower gas scale height. This strengthens the instability and makes the gas flow much more irregular at greater heights in the nonlinear state. 

\begin{figure*} 
\centering
\includegraphics[trim=0cm 0.0cm 0cm 0cm, clip=True, width=1.85\columnwidth]{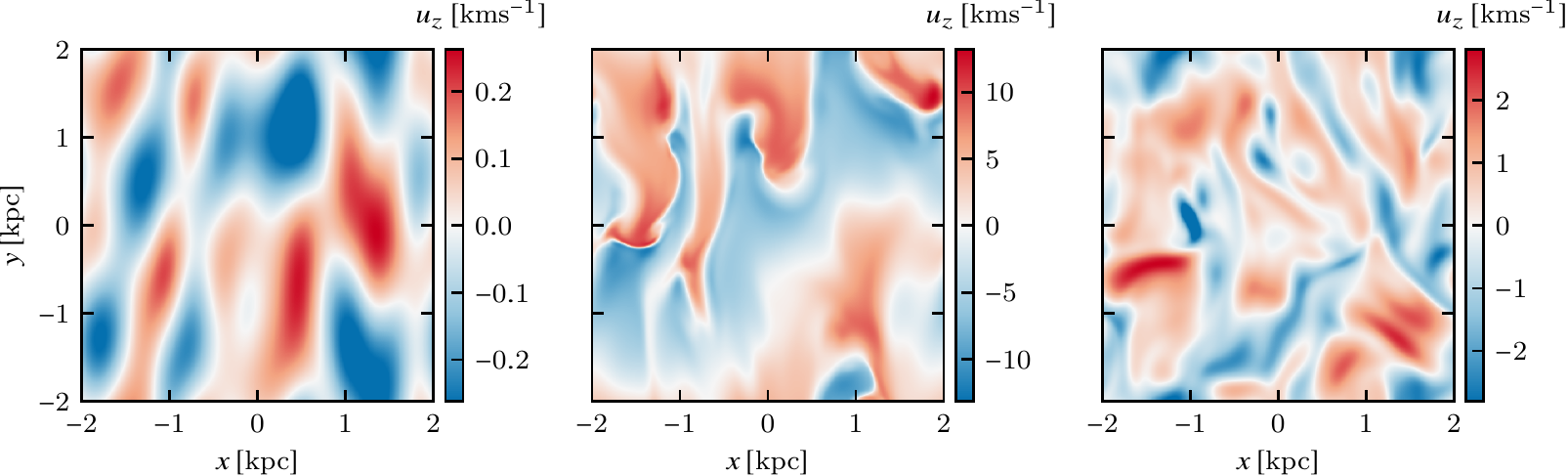}
  \begin{picture}(2,0.)(0,0)
    \put(0.08,0.52){{\sf\bf{(a)}}}
    \put(0.72,0.52){{\sf\bf{(b)}}}
    \put(1.32,0.52){{\sf\bf{(c)}}}
  \end{picture}
	\caption{The vertical velocity perturbation $u_z$ from Model Sim6 at $z=0.4\kpc$ during \textbf{(a)}~the linear stage of the instability ($t=0.3\Gyr$), \textbf{(b)}~the transitional stage ($t=0.6\Gyr$) and \textbf{(c)}~the nonlinear state ($t=0.9\Gyr$).
}
\label{fig:mbi_xy_pert}
\end{figure*}
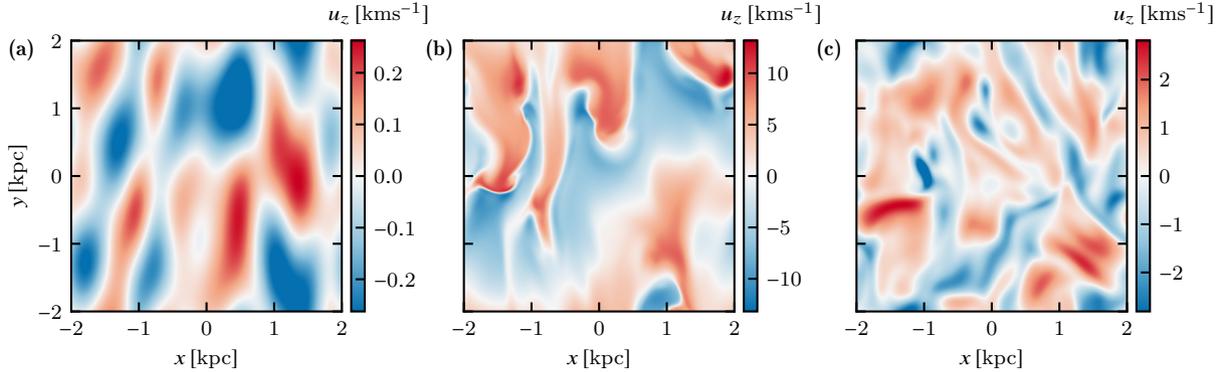

\begin{figure*}
\centering
\includegraphics[width=1.85\columnwidth]{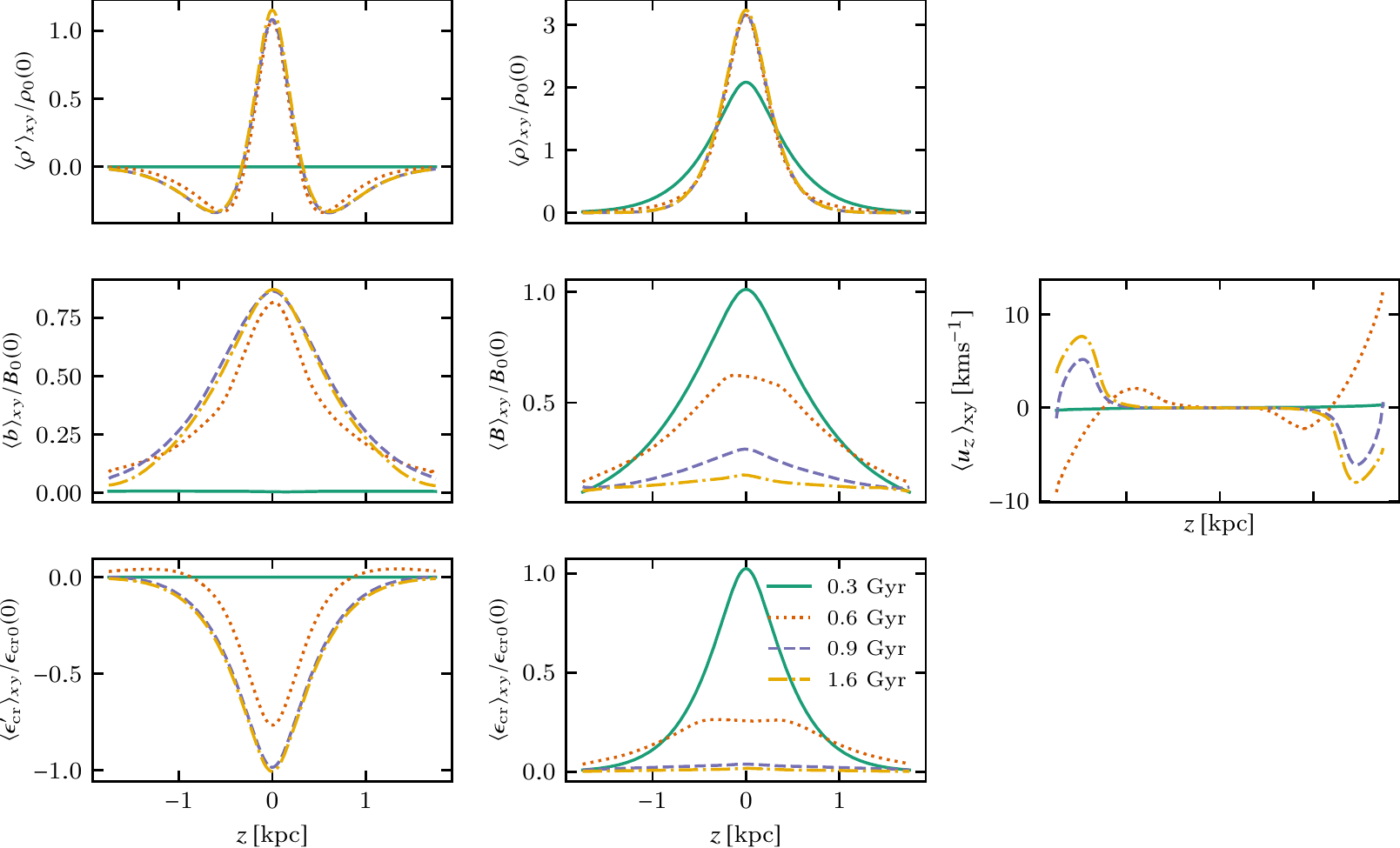}
  \begin{picture}(2,0.)(0,0)
    \put(0.07,1.130){{\sf\bf{(a)}}}
    \put(0.73,1.130){{\sf\bf{(d)}}}
    \put(0.07,0.760){{\sf\bf{(b)}}}
    \put(0.73,0.760){{\sf\bf{(e)}}}
    \put(1.37,0.760){{\sf\bf{(g)}}}
    \put(0.07,0.395){{\sf\bf{(c)}}}
    \put(0.73,0.395){{\sf\bf{(f)}}}
  \end{picture}
	\caption{
	From Model Sim6 the horizontally averaged vertical profiles of (left
	column \textbf{(a)}--\textbf{(c)}) the normalised perturbations and
	(centre  column \textbf{(d)}--\textbf{(f)}) 
	the corresponding total profiles.
	The solid, dotted, dashed and dash-dotted lines
	Timing of each profile is as indicated in the legend of \textbf{(f)}.
	respectively.
	Panel \textbf{(g)} shows the
	horizontally averaged vertical velocity normalised to the sound speed
	$\meanh{u_z}/c\sound$.
} 
\label{fig:kg89_profiles} 
\end{figure*}
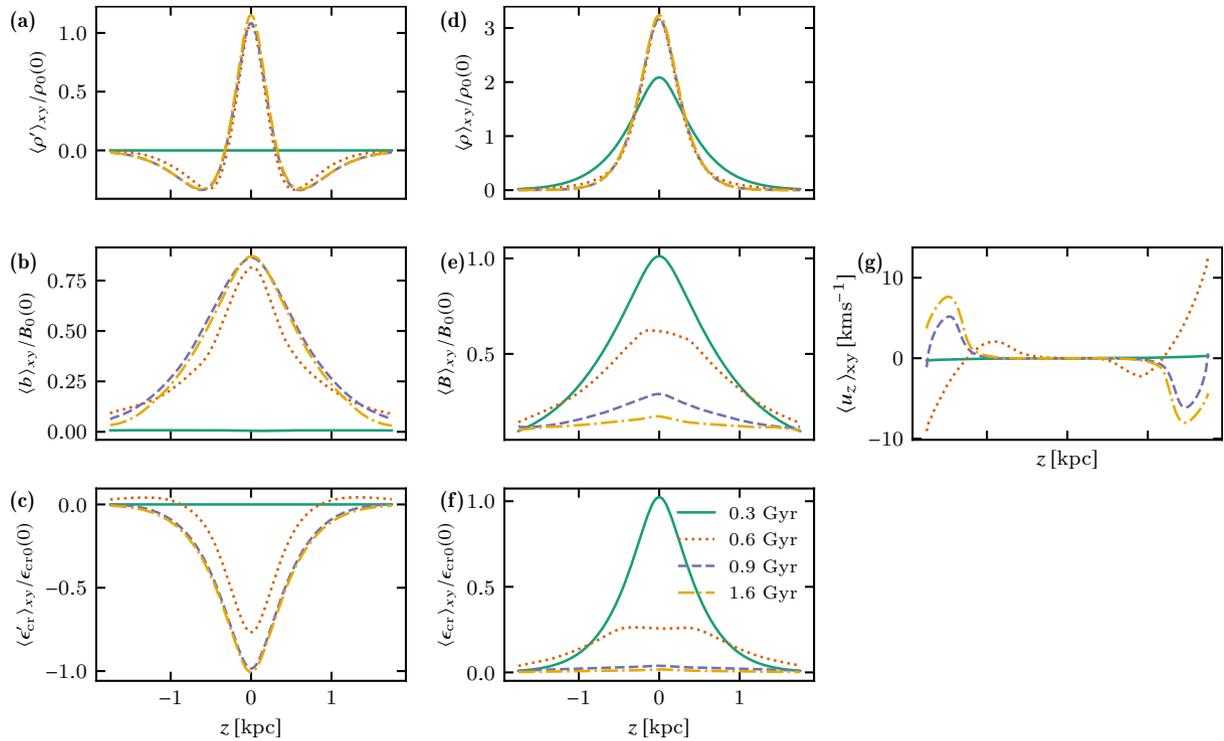 

\begin{figure}
\centering
\includegraphics[trim=0cm 0.25cm 0cm 0cm, clip=True, width=0.85\columnwidth]{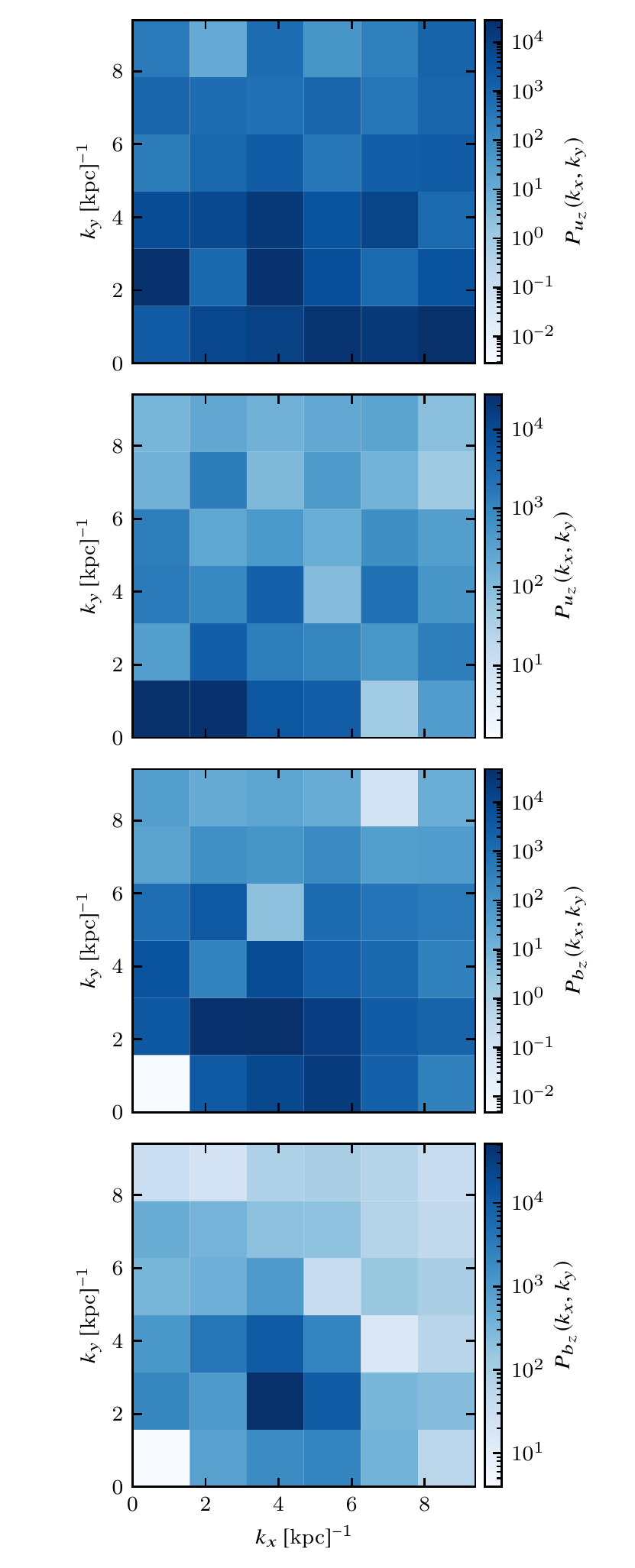}
 \begin{picture}(1.0,0)
 \put(0.16,2.070){\sf{\bf{(a)}}}
 \put(0.16,1.555){\sf{\bf{(b)}}}
 \put(0.16,1.065){\sf{\bf{(c)}}}
 \put(0.16,0.540){\sf{\bf{(d)}}}
 \end{picture}
\caption{As Fig.~\ref{fig:2psdkg89} at $|z|\leq 1.75\kpc$, but $t=0.9\Gyr$ (the
	nonlinear stage of the instability) for the perturbations
	as indicated in the colour bar labels.
	Panels {\textbf{(a)}} and {\textbf{(c)}} depict 
	Model Sim5 and {\textbf{(b)}} and {\textbf{(d)}} represent Model Sim6.
}
\label{2DN-L} 
\end{figure} 

\subsection{The nonlinear instability}\label{NLInst}

The evolution of the \rms\ velocity and magnetic fields shown in
Fig.~\ref{fig:kg89rmGR} marks three distinct stages of the instability
development. The initial stage of exponential growth is followed by a short
transitional (weakly nonlinear) period, which starts at $t=0.4\Gyr$ and lasts
for around $0.1\Gyr$ during which the growth slows down and the instability saturates
to the final steady state. The total magnetic and cosmic ray 
energy densities, which remain nearly constant throughout the linear phase,
start decaying during the transitional phase.
In the statistical steady state they retain only a few percent of the initial values. Meanwhile,
the thermal pressure in the system remains almost constant throughout the
simulation, except for a small decrease (by $4\%)$ in the nonlinear state. 

The changes in the system as the instability evolves are illustrated in Fig.~\ref{fig:mbi_xy_pert} which shows the horizontal cross-section of the vertical velocity. The pattern of the perturbations at the linear stage is quite regular and, of course, the perturbations average to zero in the horizontal planes (Fig.~\ref{fig:mbi_xy_pert}a). However, this pattern undergoes substantial modification during the transitional stage when a systematic transient inflow
develops at this altitude (Fig.~\ref{fig:mbi_xy_pert}b).
An outflow at
higher altitudes (visible in Fig.~\ref{fig:kg89_profiles}g and  discussed in
Section~\ref{VF}) is followed by a rather chaotic pattern in the nonlinear state
(Fig.~\ref{fig:mbi_xy_pert}c). The rather regular magnetic loops
characteristic of the linear instability (Fig.~\ref{fig:kg89MFNL}) evolve into
an ostensibly random pattern.
The model of \citet{Rodrigues2016}, which is evolved as far as the
very early nonlinear stage, has a similar transition to a chaotic state.

Figure~\ref{2DN-L} shows the 2D power spectra of the perturbations in the vertical velocity and magnetic field which are useful to compare with Fig.~\ref{fig:2psdkg89}. As might be expected, nonlinear effects generally broaden the power spectra, with some notable differences between Models Sim5 and Sim6. In Model Sim5, $u_z$ develops modes with small $k_y$ but large $k_x$ while the linear mode remains prominent. The maximum in the Fourier spectrum of $b_z$ shifts to higher values of $k_x$ at the same $k_y$ as in the linear stage. Meanwhile, the maxima in the power spectra of both $u_z$ and $b_z$ in Model Sim6 shift to lower $k_y$. However, the spectrum of $u_z$ is dominated by very small $k_x$ and $k_y$ whereas the spectrum of $b_z$ is maximum at $(k_x,k_y)=(3.1,1.5)\kpc^{-1}$. Altogether, the similarity between the spatial structures of the linear modes and the nonlinear solution is limited and not straightforward.

The vertical redistribution of the gas, magnetic field and cosmic rays is shown
in Fig.~\ref{fig:kg89_profiles}, which presents the horizontally averaged
deviations from the background distributions  in Panels (a)--(c) and (g) as
well as the distributions of the total gas density, magnetic field and cosmic
rays in the linear, transitional and nonlinear phases of the instability. In
the linear stage the perturbations, periodic in horizontal planes, have
vanishing horizontal averages. Non-vanishing horizontal averages 
arise only due to nonlinear effects. The perturbation in the gas density in
the transitional and nonlinear stages is positive near the midplane and
negative away from the disc: the nonlinear instability leads to a reduction in
the density scale height, so the gas disc becomes thinner.  
As a result, the mean gas density at the midplane increases from $7\times10^{-25}\g\cm^{-3}$ to
$1.08\times10^{-24}\g\cm^{-3}$ as the instability develops. The 
average energy densities of the total magnetic field and cosmic rays at the midplane
are reduced by more than $75\%$, expanding their vertical profiles. Similar
behaviour occurs in the simulations of \citet[][e.g., their
Fig.~10]{Heintz2019} which capture the weakly nonlinear, transitional stage of
the instability.

Figure~\ref{fig:kg89_profiles}g shows the vertical velocity planar averages $\langle u_z\rangle_{xy}$ at
different times. The nonlinear effects drive a systematic inflow at
$|z|\lesssim1\kpc$ and an outflow at $|z|\ga1\kpc$ in the transitional stage
which, however, is transformed into a weaker inflow at $t=0.9\Gyr$ while the
system still adjusts towards the steady state. These features are further discussed
in Section~\ref{VF}.
At $t=1.6\Gyr$, the system of Model Sim6 reaches a statistically steady state with a residual inflow with $|\meanh{u_z}|/c\sound\simeq0.5$ at $|z|\ga1\kpc$ (which, however, carries little mass). 
The magnetic field and cosmic ray energy density continue to decrease, saturating at the midplane values  $ \meanh{B(0)}/B_0(0)\approx0.16$ and  $ \meanh{\ecr(0)}/\ecri(0)\approx0.02$, while the gas density increases to $\meanh{\rho(0)}/\rho_0(0)\approx1.6$.

The magnetic buoyancy instability is driven by the vertical gradient of the magnetic field strength, and it might be expected that it would saturate via reducing the gradient to a marginal value. However, the system follows a much more dramatic path, getting rid of the magnetic field altogether. As the instability produces strong vertical magnetic perturbations, the cosmic rays are channelled out from the disc and diffuse along the magnetic field at a high rate. The instability results in wide distributions of both magnetic field and cosmic rays enveloping a relatively thin thermal gas disc. A similar kind of evolution also occurs in the simulations of \citet{Heintz2019} and \citet{GPPS22}. However, the form of the nonlinear state is sensitive to such features of the system as rotation, and to the mechanism maintaining the unstable magnetic field in the disc. As we show in \citet{TSGSS22}, the instability in a rotating system can lead even to the reversal of the magnetic field in the disc.

\section{Discussion}\label{discussion}
In this section we discuss the implications of our results for the dynamics of the interstellar medium and describe the force balance in the system.

\subsection{Cross-correlation between energy densities}\label{CCBED}
Understanding the relative spatial distributions of the gas, magnetic field and cosmic rays in the interstellar medium is crucial for the interpretation of the observations in the radio and other wavelength ranges. In particular, the assumption of a tight correlation between cosmic rays and magnetic fields (e.g., the energy equipartition) is routinely used in the interpretations of synchrotron observations \citep[][and references therein]{SeBe19}. The correlation or anti-correlation between the thermal electron density and magnetic field can affect significantly the Faraday rotation \citep{BSSW03}.

\begin{table}
\caption{The Pearson correlation coefficients $r$ between  various energy densities in the statistically steady state at $t=1.6\Gyr$ presented as $a,b$, where $a$ and $b$ refer to the altitudes $z=0.5$ and $1.5\kpc$, respectively.}
\centering
\begin{tabular}{@{}ccccc@{}}
\hline
                           &$\epsilon_\text{th}^\prime$    &$\ecr^\prime$                  &$\epsilon\m^\prime$            &$\epsilon_\text{k}^\prime$\\
\hline 
$\epsilon_\text{th}^\prime$&$\phantom{-00}1,\phantom{-00}1$&$-0.70,-0.64$                  &$-0.80,-0.17$                  &$\phantom{-}0.17,-0.22$          \\
$\ecr^\prime$              &                               &$\phantom{-00}1,\phantom{-00}1$&$\phantom{-}0.58,-0.46$        &$-0.2\phantom{0},-0.47$          \\
$\epsilon\m^\prime$        &                               &                               &$\phantom{-00}1,\phantom{-00}1$&$\phantom{-}0.01,\phantom{-}0.27$\\
$\epsilon_\text{k}^\prime$ &                               &                               &                               &$\phantom{-00}1,\phantom{-00}1$  \\
\hline
    \end{tabular}
    \label{tab:corrcoef_full}
\end{table}

Table~\ref{tab:corrcoef_full} presents the Pearson cross-correlation coefficient $r$ between the fluctuations in the energy densities $\epsilon_i'=\epsilon_i-\langle \epsilon_i\rangle_{xy}$, with $\langle\cdots\rangle_{xy}$ for the horizontal average and $i = \text{th, cr, m, k}$ for the thermal, cosmic-ray, magnetic and kinetic ($\epsilon_\text{k}=\tfrac12\rho u^2$) energy densities, in the nonlinear stage: 
\begin{align}\label{e_flucs}
\epsilon_{\rm m}' &= \frac{B^2}{8\pi}  - \left\langle \frac{B^2}{8\pi}\right\rangle_{xy}, & \ecr'&= \ecr - \langle \ecr \rangle_{xy}\,,\\\nonumber
\epsilon_{\rm th}' &= c\sound^2\left( \rho - \langle \rho \rangle_{xy}\right), & \epsilon_{\rm k}'&=\tfrac{1}{2}\rho u^2-\left\langle\tfrac{1}{2}\rho u^2\right\rangle_{xy}\,.
\end{align}

\begin{figure}
    \centering
    \includegraphics[width=0.8\columnwidth]{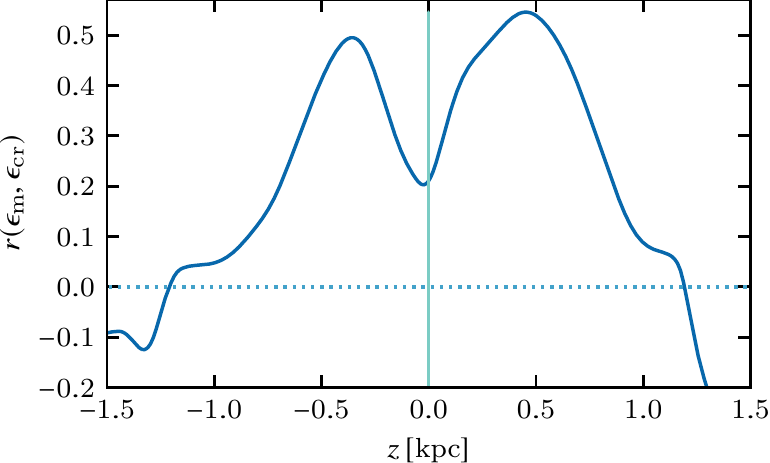}
    \caption{The vertical variation of the Pearson cross-correlation coefficient between the  magnetic  ($\epsilon\m$) and cosmic ray ($\ecr$) energy densities in Model Sim6 averaged over the time interval $1.2\leq t\leq1.7\Gyr$ with the cadence $\Delta t=0.1\Gyr$.}
    \label{fig:bcr_corrz}
\end{figure}

The relations between these variables are different near the midplane and at a higher altitude, so both are presented (the results at $z<0$ are similar). The fluctuations in thermal energy density are significantly anti-correlated with those in the magnetic and cosmic ray energy densities near the midplane and with the cosmic ray energy fluctuations at the higher altitude.
This appears to reflect the horizontal pressure balance where the total average pressure is independent of $x$ and $y$. The kinetic energy density shows no significant correlation with any other variable. The distribution of the cosmic rays is only weakly correlated with that of the magnetic field near the midplane ($r=0.58$) but not away from it ($r=-0.46$). Figure~\ref{fig:bcr_corrz} shows the cross-correlation coefficient of the fluctuations in the magnetic field and cosmic ray energy densities at various heights in the nonlinear stage of Model Sim6.
The cross-correlation coefficient of these two quantities increases with $|z|$ reaching a maximum at $|z|\simeq0.5\kpc$ and then decreases to become negative (anti-correlation) at $|z|\ga1\kpc$.

To clarify the cause of the correlations, consider the magnetic field and cosmic rays. The magnetic field and cosmic ray energy density can be decomposed into their horizontal averages and fluctuations, $\vec{B}=\vec{\mean{B}}+\vec{b}$ and $\ecr=\mean{\epsilon}+\epsilon'$, where $\vec{\mean{B}}=\meanh{\vec{B}}$ and $\mean{\epsilon}=\meanh{\ecr}$, such that $\Meanh{\mean{B}}=\mean{B}$, $\meanh{\vec{b}}=0$, $\meanh{\mean{\epsilon}}=\mean{\epsilon}$ and $\meanh{\epsilon'}=0$.
The correlation coefficient of the total magnetic and cosmic ray energies follows as
\begin{equation}\label{rBEdef}
r(\epsilon\m,\ecr)=\frac{\Meanh{\left(B^2-\meanh{B^2}\right)\left(\ecr-\mean{\epsilon}\right)}}{\sigma_\text{m}\sigma_\epsilon}\,,
\end{equation}
where $\sigma_m^2=\meanh{B^4}-\meanh{B^2}^2$ and $\sigma_\epsilon^2=\meanh{(\epsilon')^2}$ are the variances of the magnetic (the factor $8\pi$ can be omitted here) and cosmic ray energy densities. Since $B^2=\mean{B}^2 +2\vec{\mean{B}}\cdot\vec{b} + b^2$, $\meanh{B^2}=\mean{B}^2+\meanh{b^2}$ and $\Meanh{\meanh{b^2}\epsilon'}= \meanh{b^2}\,\meanh{\epsilon'}=0$, we have
\begin{equation}\label{rBE}
r(\epsilon\m,\ecr)=\frac{\Meanh{\left(2\vec{\mean{B}}\cdot\vec{b}+b^2\right)\epsilon'}}{\sigma_\text{m}\sigma_\epsilon}\,.
\end{equation}
Both terms contribute similarly to the correlation coefficient: $r(b^2,\ecr)=0.23$ and $r(\vec{B}\cdot\vec{b},\ecr)=0.43$ at $z=0.5\kpc$ and $-0.48$ and $-0.18$ at $z=1.5\kpc$, respectively.

The variance of the magnetic energy density can be represented as
\begin{equation}\label{sB2}
\sigma\m^2=4\Meanh{(\vec{\mean{B}}\cdot\vec{b})^2} +\meanh{b^4}-\meanh{b^2}^2\,,
\end{equation}
since $\meanh{b^2\vec{\mean{B}}\cdot\vec{b}}=0$ for symmetric probability distributions of the Cartesian components of $\vec{b}$.

\begin{figure*}
    \centering
    \includegraphics[width=1.85\columnwidth]{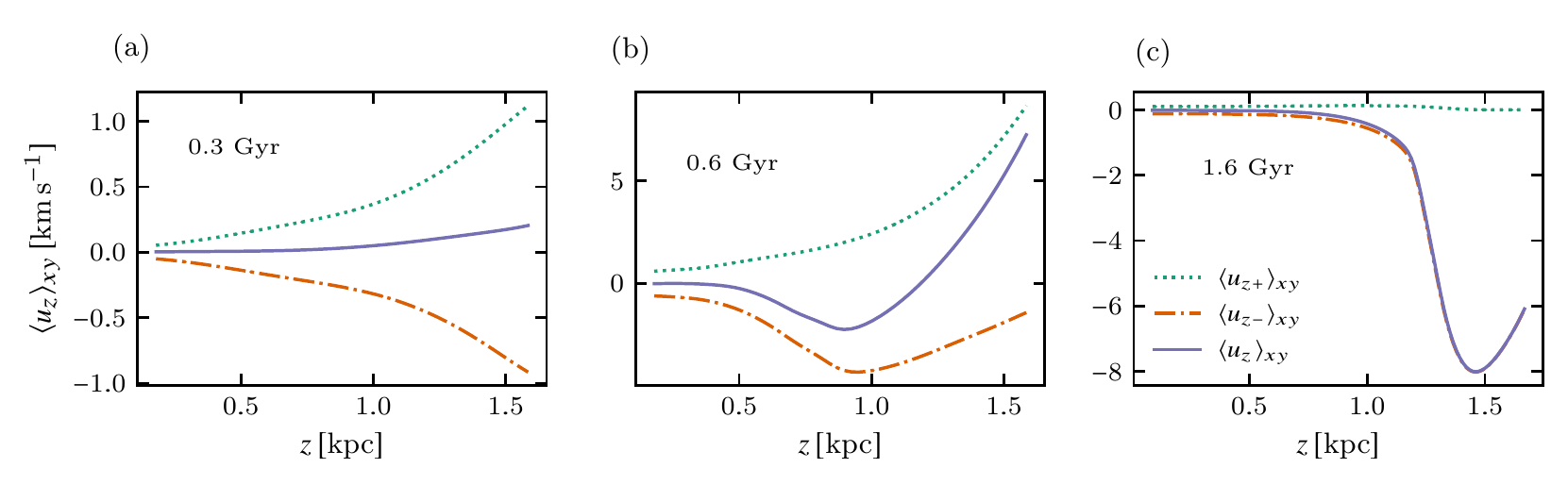}
    \caption{The variation of the horizontally averaged vertical velocity component, $u_z$ (solid lines) at $z>0$ in Model Sim6 at \textbf{(a)}~$t=0.3\Gyr$, \textbf{(b)}~$0.6\Gyr$ and \textbf{(c)}~$1.6\Gyr$. The horizontally averaged outflow (inflow) speed is shown dotted (dash-dotted) while the horizontally averaged velocity, their sum, is shown solid. 
    }
    \label{fig:kg89uz}
\end{figure*}

To provide an illustration, if the magnetic field fluctuations are isotropic and the Cartesian component $b_i$ of $\vec{b}$ are statistically independent, each having the standard deviation $\sigma_i$, we have $\meanh{b^2}=3\sigma_i^2$ and $\meanh{b_i^4}=3\sigma_i^4$, so that $\meanh{b^4}=\Meanh{\left(\sum_{i=1}^3 b_i^2\right)^2}=15\sigma_i^2$ and $\meanh{b^4}-\meanh{b^2}^2=6\sigma_i^2$.

As discussed by  \citet{BSSW03} \citep[see also Sect.~13.2 of][where typographical errors of the original publication are corrected]{ShSu22}, the anti-correlation of the total magnetic field strength and thermal gas density can significantly affect the interpretation of the Faraday rotation observations in terms of the magnetic field strength, leading to underestimated field strength when the two variables are assumed to be uncorrelated. The weak correlation between the total magnetic and cosmic ray energy densities near the midplane, $r\approx0.6$, and a similarly weak anti-correlation at a higher altitude, $r\approx-0.5$, are inconsistent with the assumption of the \textit{local} energy equipartition between magnetic fields and cosmic rays at the scales of the fluctuations produced by the Parker instability, of order $1\kpc$ and less.

The energy equipartition between cosmic rays and magnetic fields is often justified by arguing that cosmic rays can accumulate in the galactic disc until they achieve the energy density comparable with that of the magnetic field. Then their pressure can open up magnetic lines locally and they can be released from the galaxy, after which their pressure decreases down to a level controlled by the magnetic field. Such a self-regulation is argued to lead to their energy equipartition. Our model contains all the elements of these processes in their full nonlinear implementation, and yet no equipartition occurs. In particular, vertical magnetic fields that facilitate the vertical diffusive transport of the cosmic rays are produced by the magnetic buoyancy without any need for the cosmic ray pressure to affect the structure of the magnetic field (even though cosmic rays can enhance the magnetic buoyancy instability). 

\subsection{Vertical flows}\label{VF}
The redistribution of the gas, magnetic field and cosmic rays in the unstable system involves systematic vertical flows. As discussed above, the vertical velocity averaged over a horizontal plane necessarily vanishes in the linear stage of the instability but, remarkably, systematic flows emerge in the transitional stage as a nonlinear effect. Figure~\ref{fig:kg89uz} clarifies the balance between the inflow ($zu_z<0$) and outflow ($zu_z>0$) as the instability develops presenting, for $0\leq z \leq 1.75\kpc$ (the picture at $z<0$ is similar), separately the vertical velocity averaged over those regions in the $(x,y)$-planes where $u_z>0$ and $u_z<0$, denoted $\langle u_z\rangle_+$ and $\langle u_z\rangle_-$, respectively. In the linear stage, $t=0.3\Gyr$, the outflows balance the inflows and the horizontally averaged vertical velocity $\langle u_z\rangle_{xy}= \langle u_z\rangle_- + \langle u_z\rangle_+$ is negligible. As the nonlinear effects become stronger, at $t=0.6\Gyr$, a systematic inflow develops around $z=1\kpc$ and outflow with the maximum outflow speed of $7\km\s^{-1}$ is maintained at higher altitudes. At a later time, $t=0.9\Gyr$, when the nonlinear effects are still stronger but the system continues evolving, an inflow dominates with $-4\km \s^{-1}$ at $z=1.5\kpc$. In the advanced nonlinear stage at $t=1.6\Gyr$, the \rms\ inflow speed saturates at $-8\kms$ at $z=1.5\kpc$.

\begin{figure} \centering
\includegraphics[width = 0.85\columnwidth]{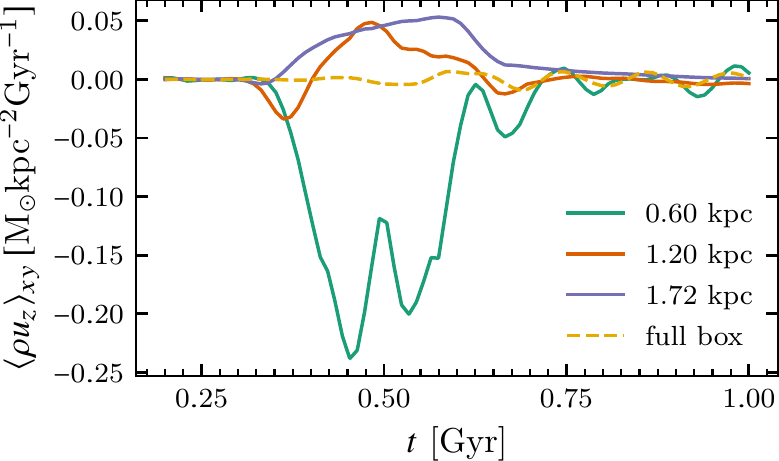}
\caption{The evolution of the mass flux in Model Sim6
at various heights above the midplane as specified in the legend.
} 
\label{fig:kg89_momflux_z} 
\end{figure} 

\begin{figure*}
 \centering
\includegraphics[width=1.85\columnwidth]{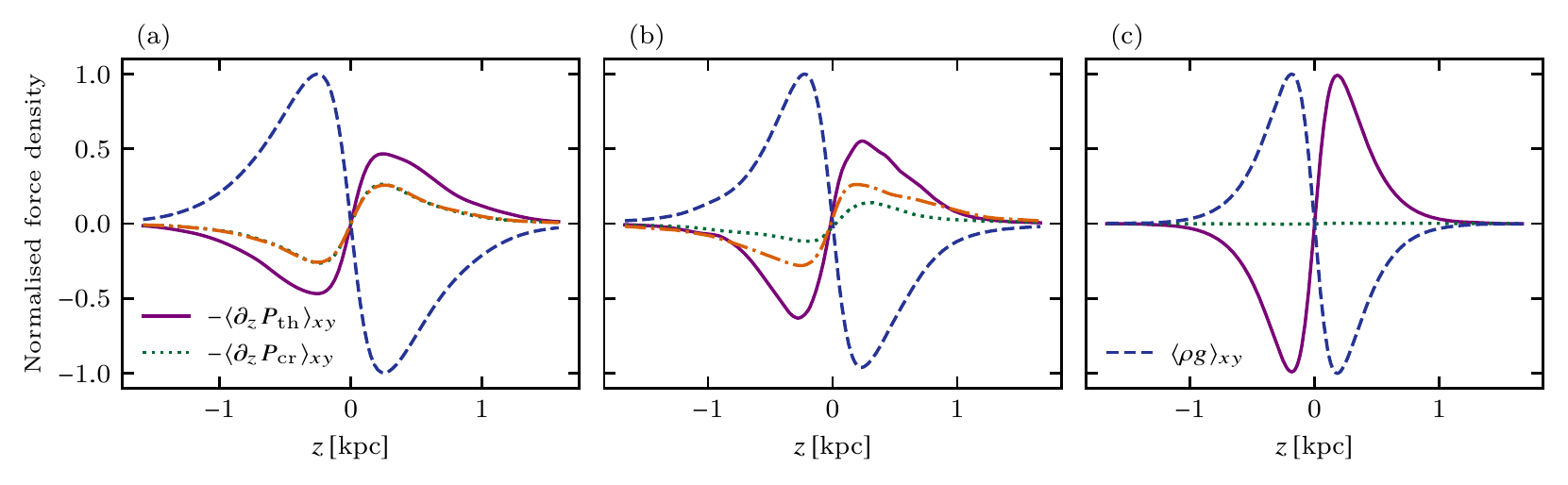}
\caption{The vertical variation of the horizontally averaged forces normalised to the maximum magnitude of the gravitational force (dashed) at different stages of the development of the instability in Model Sim6: \textbf{(a)}~$t=0.3\Gyr$ (linear instability), \textbf{(b)}~$0.5\Gyr$ (transitional; the magnetic field and cosmic rays have identical mean pressure distributions) and \textbf{(c)}~$1.6\Gyr$ (nonlinear state; the mean pressure gradients of the magnetic field and cosmic rays are negligible). The gas pressure gradient (solid) is stronger than the pressure gradients of the magnetic field (dash-dotted) and cosmic rays (dotted), and this dominance is extreme in the nonlinear state.} 
\label{fig:kg89_forcebalance}
\end{figure*}

The fact that the Parker instability can drive systematic vertical flows can be of significance for the galactic evolution and regulation of star formation in the disc. Figure~\ref{fig:kg89_momflux_z} shows the evolution of the mass
flux at different heights. In the transitional phase ($0.4 < t < 0.5\Gyr$), the system develops a strong inflow near the midplane. 
During this period, the gas is redistributed into a thinner layer.
As the system evolves into the nonlinear phase ($ t > 0.5\Gyr$), the mass flux reduces, through decaying oscillations to become negligible in the late nonlinear stage. The outflow at higher altitudes discussed above involves dilute gas and does not transport any significant gas mass in this model.

\subsection{Force balance}\label{FB}
Figure~\ref{fig:kg89_forcebalance} shows the horizontally averaged values of the thermal, magnetic, cosmic ray pressure gradients (the averaged magnetic tension force is negligible at all times) as well as the gravitational force at three different stages of the evolution. During the linear stage (Fig.~\ref{fig:kg89_forcebalance}a) the horizontal averages of the magnetic field and cosmic rays vanish, and the average force balance is exactly the same as in the background distributions. From the transitional phase to the nonlinear state (Fig.~\ref{fig:kg89_forcebalance}b,c, respectively), the contributions of the magnetic and cosmic ray pressures decrease systematically as their scale heights increase, whereas the gradient of the thermal pressure increases because the gas scale height decreases (see Fig.~\ref{fig:kg89_profiles}).

\begin{figure}
\centering
\includegraphics[width =0.85\columnwidth]{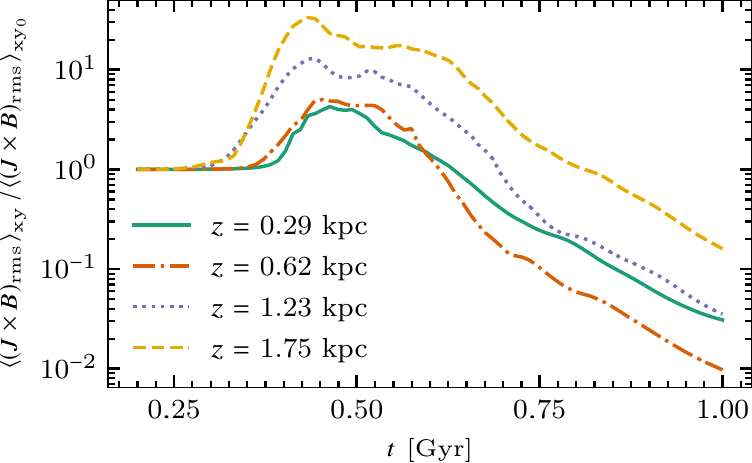}
\caption{Horizontally averaged \rms\ Lorentz force of Model Sim6 
	(normalised by its midplane magnitude) at distance above the midplane as listed in the
	legend. $\vec{J}$ is electric current density.
}
\label{fig:jxb_evolve}
\end{figure}

\begin{figure*}
\centering
\includegraphics[width=1.65\columnwidth]{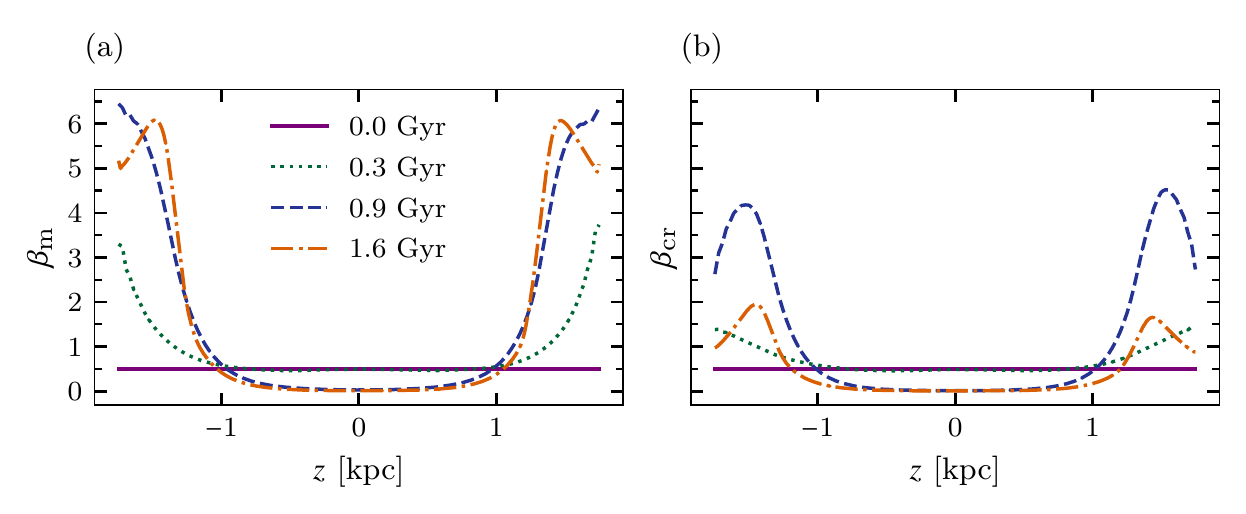}
\caption{
	From Model Sim6 the horizontally averaged ratios to the thermal
	pressure, see Eq.~\eqref{eq:pressures}, of \textbf{(a)}  magnetic
	pressure $\beta\m$ and \textbf{(b)}  cosmic ray pressure $\beta\cra$.
	Vertical distributions are shown at times
	specified in the legend, which represent the initial state, linear,
	transitional and nonlinear stages of the instability, respectively.
} 
\label{fig:beta_gasbeta_cr_ratio}
\end{figure*}

To clarify the nature of the gas outflow prominent during the transitional stage of the instability, we show in Fig.~\ref{fig:jxb_evolve} the evolution
of the magnitude of the horizontally averaged Lorentz force 
per unit volume at different heights. 
During the linear phase of the instability, the average magnetic force remains equal to that in the background (initial) state but it increases sharply during the transitional phase as significant vertical magnetic fields emerge. This increase is stronger at high altitudes (by a factor of 30). Since the gas density decreases with $|z|$, the increase in the force per unit mass is even stronger, and the magnetic force is clearly the driver of the systematic gas outflow in the transitional state (Section~\ref{VF}). As the instability saturates, the Lorentz force decays leaving for the thermal pressure gradient alone to balance the gravitational force.

In the initial state, the magnetic and cosmic ray pressures vary with $z$ exactly as the thermal pressure, so both $\beta\m$ and $\beta\cra$ of  Eq.~\eqref{eq:pressures} are independent of $z$. Analytical analyses of the linear Parker instability usually rely on the assumption that both $\beta\m$ and $\beta\cra$ remain independent of $z$ as the instability develops. Figure~\ref{fig:beta_gasbeta_cr_ratio} shows the variation with $z$ of the horizontally averaged pressure ratios. In the linear stage, both ratios remain unchanged at $|z|\lesssim1\kpc$, which justifies the assumption of the analytical studies, but increase significantly at $|z|\gtrsim1\kpc$, especially $\beta\m$. As the system evolves into the linear and nonlinear stages, both ratios are reduced to negligible values at $z\lesssim1\kpc$  due to the increase in the thermal pressure and the reduction in magnetic and cosmic rays pressures near the midplane caused by the redistribution of the gas towards the midplane and escape of the magnetic field and cosmic rays to greater heights. As a result, the outer layers, $|z|\gtrsim1\kpc$, are strongly dominated by the magnetic field and cosmic rays -- and yet the magnetic force and the gradient of the cosmic pressure are both negligible in comparison with the thermal pressure gradient in the nonlinear state.

\subsection{Sensitivity to parameters}\label{StP}

\begin{figure}
    \centering
\includegraphics[trim=0cm 0.2cm 0cm 0.2cm, clip=True, width=0.85\columnwidth]{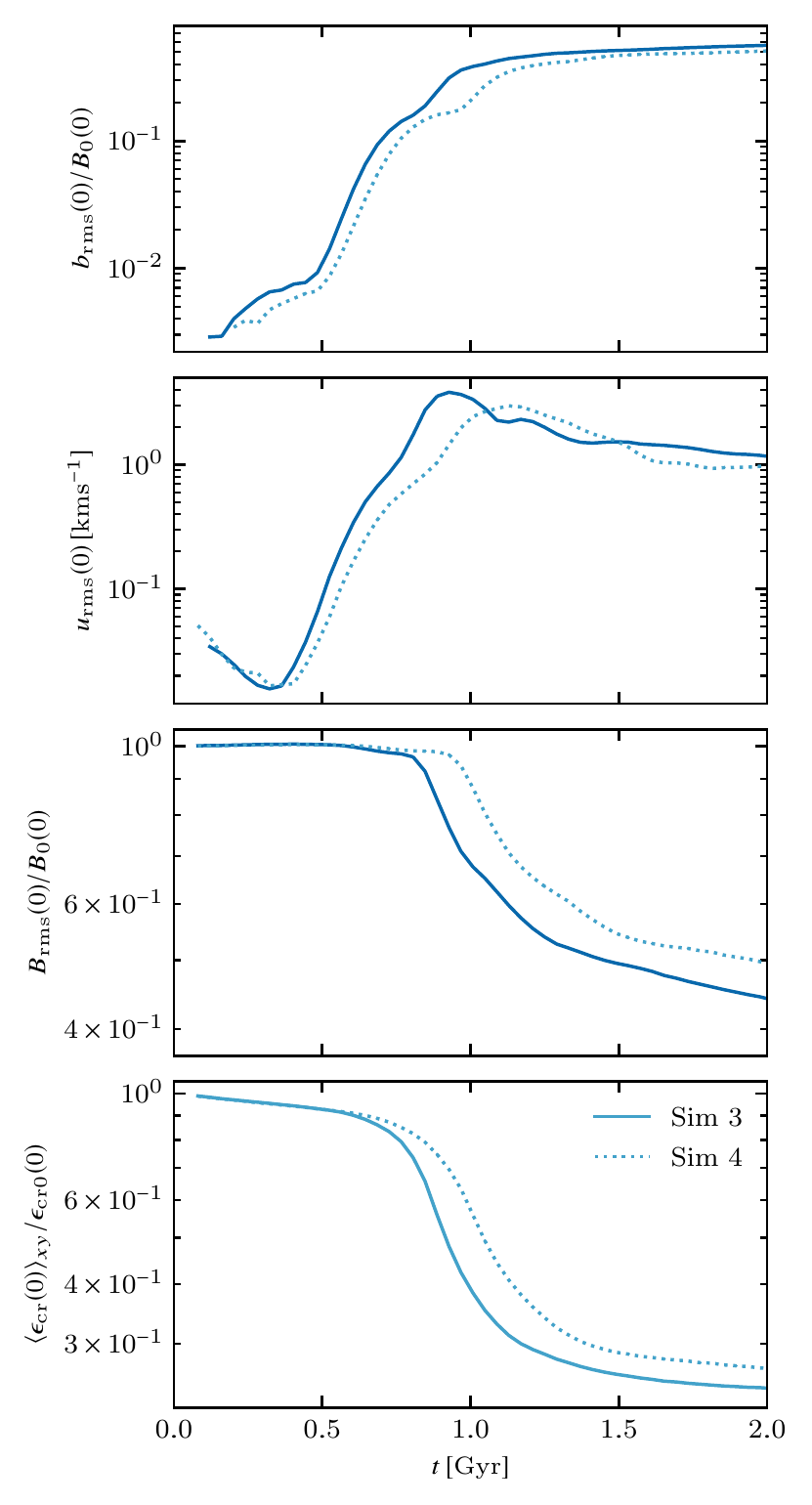}
 \begin{picture}(1,0)
 \put(0.085,1.535){\texttt{(a)}}
 \put(0.085,1.185){\texttt{(b)}}
 \put(0.085,0.795){\texttt{(c)}}
 \put(0.085,0.425){\texttt{(d)}}
 \end{picture}
    \caption{As Fig.~\ref{fig:kg89rmGR} but for Models Sim3
    (solid) and Sim4 (dotted) where the effective density scale height is $ h= 1\kpc$ in the background state.    }
    \label{fig:convhtestrms}
\end{figure}

The instability is sensitive to a wide range of parameters including the scale height of the unstable magnetic field, the ratios of the magnetic and cosmic ray pressures to the thermal pressure, $\beta\m$ and $\beta\cra$, and the form of the gravity profile $g(z)$. In this section we discuss the effects of the key parameters and assumptions, with emphasis on the nonlinear states. The models Sim1--Sim4 have the gravity profile \eqref{gtanh} while Sim5--Sim9 use $g(z)$ from equation~\eqref{KG89grav}. To assess the role of the cosmic ray propagation governed by equation~\eqref{ecr}, we also consider a model where the only effect of the cosmic rays is to contribute to the total pressure, assuming that this contribution is equal to the magnetic pressure (thus, just doubling the magnetic pressure in such a model). Parameters of the models discussed below can be found in Table~\ref{tab:my_label}.

\subsubsection{Gas scale height and nonthermal pressures}\label{SHaPR}
The purpose of Models Sim1--Sim4 is to explore the role of the gas (and magnetic field) scale heights and nonthermal pressures in the background state. The initial (background) state in Models Sim3 and Sim4 has a relatively large gas scale height $h=1\kpc$ (and a correspondingly larger scale height of the background magnetic field) and lower magnetic and cosmic ray pressures than in the reference model Sim2 (with the same gravity profile) while Model Sim1 represents the case of the pure magnetic buoyancy instability ($\beta\cra=0$). The adopted values of $\beta\m$, $\betacr$ and $h$ require a higher sound speed in these models \citep[see equation~(9) of][]{Rodrigues2016}. In this sense, these models allow for the presence of the hot interstellar gas. 

Figure \ref{fig:convhtestrms} shows the evolution of the magnetic and cosmic ray energy densities for these two systems -- it is useful to compare it with Fig.~\ref{fig:kg89rmGR}. As expected, the instability is weaker (lower growth rate of the perturbations) in Models Sim3 and Sim4 than in Sim2. The system Sim3 ($\beta\m=\betacr=0.5$) has a higher growth rate than Sim4 ($\beta\m=\betacr=0.25$) and, in the transitional stage ($1\lesssim t\lesssim2\Gyr$) loses the magnetic field and cosmic rays faster. The amount of the magnetic field and cosmic ray energies lost in the transitional stage is also slightly smaller in the system with weaker instability.

\subsubsection{Gravity profile}\label{GrPr}


\begin{figure}
\centering
\includegraphics[trim=0cm 0.25cm 0cm 0cm, clip=True, width=0.85\columnwidth]{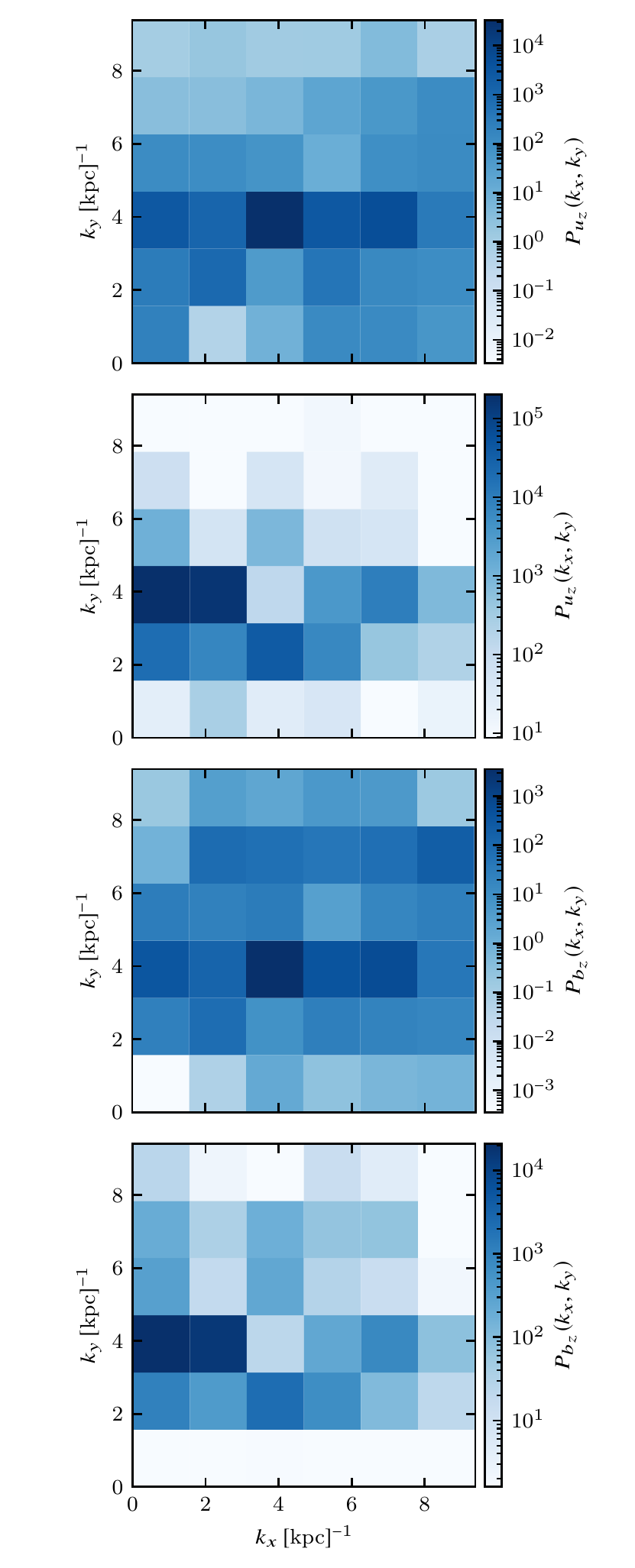}
 \begin{picture}(1.0,0)
 \put(0.16,2.070){\sf{\bf{(a)}}}
 \put(0.16,1.555){\sf{\bf{(b)}}}
 \put(0.16,1.065){\sf{\bf{(c)}}}
 \put(0.16,0.540){\sf{\bf{(d)}}}
 \end{picture}
\caption{As Fig.~\ref{fig:2psdkg89}, but at $t=0.3\Gyr$ (the
	linear stage of the instability) and for $|z|\leq 1.75\kpc$ for the perturbations
	as indicated in the colour bar labels.
	However, panels {\textbf{(a)}} and {\textbf{(c)}} here depict 
	Model Sim2, while {\textbf{(b)}} and {\textbf{(d)}} again depict Model Sim6.
}
\label{fig:Fourier2dxykg89_grav}
\end{figure}

\begin{figure}
    \centering
\includegraphics[trim=0cm 0.2cm 0cm 0.2cm, clip=True, width=0.85\columnwidth]{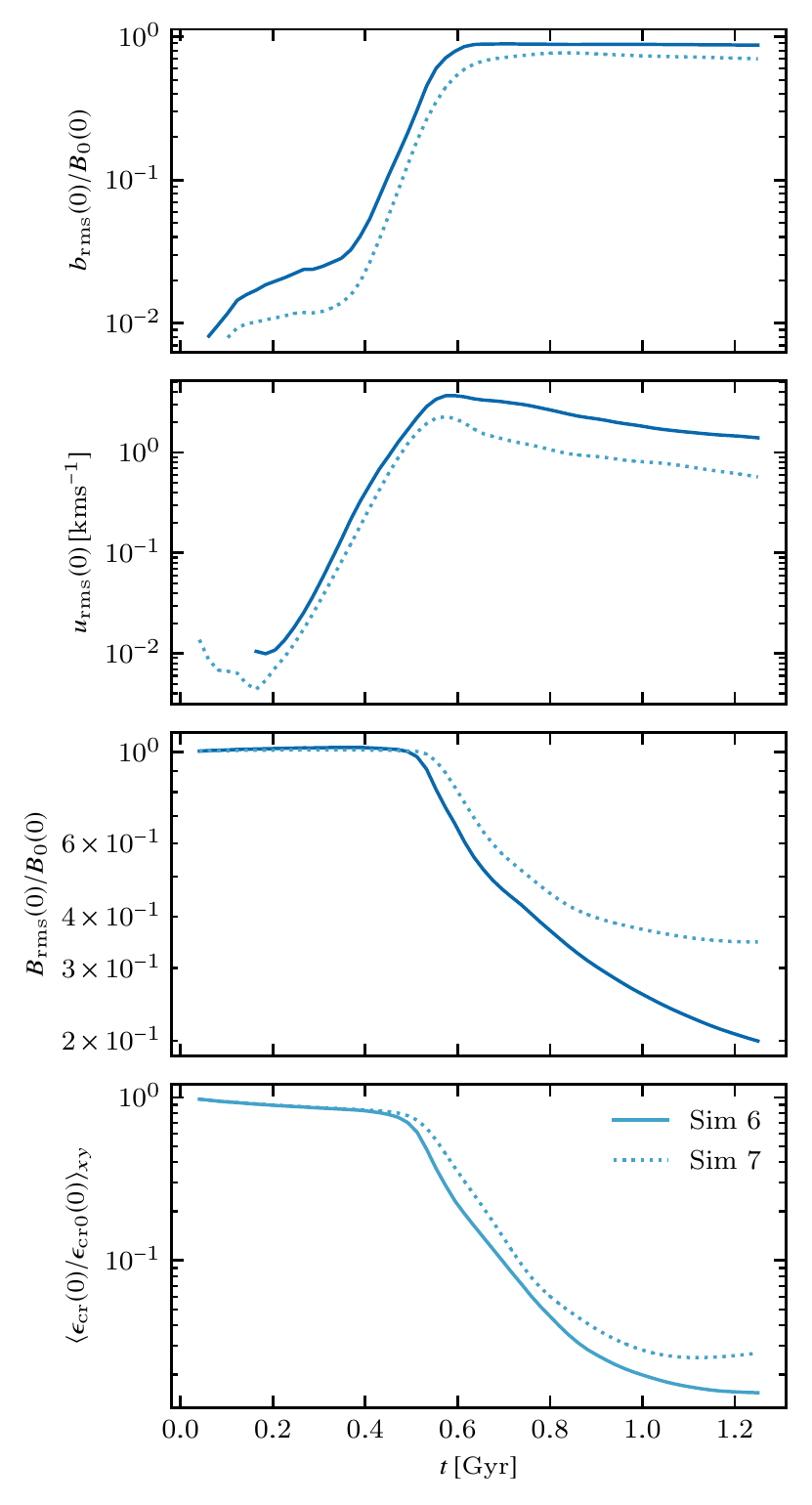}
 \begin{picture}(1,0)
 \put(0.085,1.535){\texttt{(a)}}
 \put(0.085,1.185){\texttt{(b)}}
 \put(0.085,0.795){\texttt{(c)}}
 \put(0.085,0.425){\texttt{(d)}}
 \end{picture}
    \caption{As Fig.~\ref{fig:kg89rmGR} but for Models Sim6 (solid) and Sim7 (dotted); both the viscosity and magnetic diffusivity are larger in the latter model.}
    \label{fig:convetatestrms}
\end{figure}

The difference between the gravity profiles \eqref{gtanh} and \eqref{KG89grav} is not strong and yet it can be of a physical significance since the gravity fields vary in magnitude and form between different galaxies and between different locations within a specific galaxy.  In the linear stage, the development of the instability mostly depends on the system properties at $|z|\lesssim h$, where $h=0.3\text{--}0.5\kpc$ in most of the models considered here, so it might by reasonable to expect the modes of the linear instability to be broadly similar under both gravity profiles but it is useful to identify any subtler changes. In 
Models Sim2 and Sim6, which only differ in the gravity profiles, the growth rate of the instability is about $25\Gyr^{-1}$.

In Fig.~\ref{fig:Fourier2dxykg89_grav}, we compare the 2D power spectra of the vertical velocity $u_z$ and magnetic field $b_z$ in the linear instability phase for Models Sim2 and Sim6; the higher value of $h$ in Model Sim2 reflects the somewhat weaker gravity. As also shown in Table~\ref{tab:my_label}, both models have similar wave number $k_y$ of the most unstable mode although the spectrum in $k_y$ around the maximum at $k_y \approx3 \kpc^{-1}$
is wider in Model Sim2 (suggesting that this mode is less dominant). Furthermore, the mode structures in $x$ and $z$ differ significantly: \eqref{gtanh} leads to significantly larger values of $k_x$ and a weaker variation of the solution with $z$. The nonlinear state can be expected to be more sensitive to the system properties at greater distances from the midplane where the two gravity profiles differ stronger \citep[see also][]{LBS17}. However, this difference within our simulation domain is not strong enough to affect, in particular, the vertical flows. In Models Sim2 and Sim6, the outflow speeds during the transitional stage at $t = 0.6\Gyr$ are $\meanh{u_z} = 7 \kms$ and $12\kms$ at $z= 1.5\kpc$. Similarly, the maximum magnitude of the downward velocity in the late nonlinear stage at $t = 1.25\Gyr$ is $5\kms$ and $8\kms$, respectively. 

\begin{table*}
\caption{The wave numbers $k_x$, $k_y$ and $k_z$ of the mode that grows most rapidly in simulations with the same parameters as Model Sim6 but with the computational domain of the lengths $L_x$, $L_y$ and $L_z$ along the $x$-, $y$- and $z$-axes, and the corresponding numerical resolutions $\Delta x$, $\Delta y$ and $\Delta z$, where the smallest admissible wave numbers are $2\pi/L_x$, $2\pi/L_y$ and $2\pi/L_z$ and the largest are $2\pi/\Delta x$, $2\pi/\Delta y$ and $2\pi/\Delta z$,  respectively. Each row presents a separate simulation.
}
\centering
\begin{tabular}{cccccccccccccc}
\hline
$L_x$  &$\Delta x$ &$2\pi/L_x$     &$k_x$   &   &$L_y$ &$\Delta y$ &$2\pi/L_y$  &$k_y$ &  &$L_z$ &$\Delta z$ &$2\pi/L_z$  &$k_z$\\[0pt] 
[kpc]  &[pc] &[kpc$^{-1}$]   &[kpc$^{-1}$]  & &[kpc] &[pc] &[kpc$^{-1}$]   &[kpc$^{-1}$] & &[kpc] &[pc] &[kpc$^{-1}$]   &[kpc$^{-1}$]\\[3pt]
\cline{1-4} \cline{6-9}    \cline{11-14}\\[-6pt]
8       &31  &0.8            &1.6      &      &7     &13   &0.9            &2.8&&3.5   &13 &1.8 &3.6\\
20      &52  & 0.3            &1.6      &      &24    &46   &0.3            &2.8 &&3.5   &13 &1.8 &3.6\\
\hline
    \end{tabular}
    \label{LaDo}
\end{table*}
\begin{figure*}
\centering
\includegraphics[width=1.95\columnwidth]{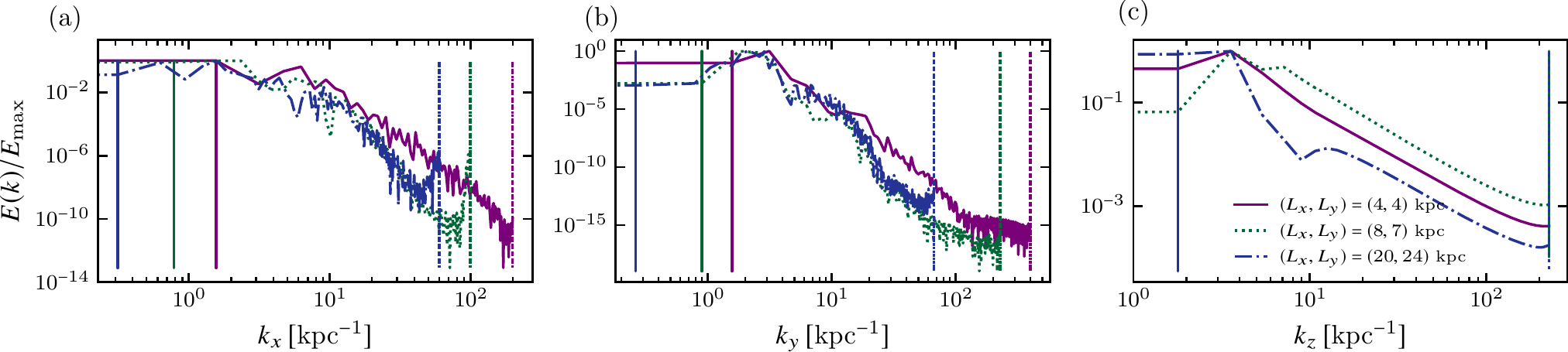}
\caption{The normalised power spectra \textbf{(a)}~$E(k_x)$, \textbf{(b)}~$E(k_y)$ and \textbf{(c)}~$E(k_z)$ of $u_z$
	(in $\rm kpc^2km^2\,s^{-2}$) for models of with domain sizes listed in the legend. Otherwise 
	parameters are as in Model Sim6. Spectra are time-averaged, sampled every $10\,\rm Myr$ for $0.25\leq t\leq0.30\Gyr$. 
	The vertical lines of matching colour for each model mark the minimum (solid)
	and maximum (dotted) admissible
	wave numbers.}
    \label{PoSp}
\end{figure*}

\subsubsection{The size of the computational domain}\label{SCD}
The size of the computational domain needs to be large enough to accommodate the most rapidly growing mode. Because of the periodic boundary conditions in $x$ and $y$, only a discrete set of modes can be excited in the numerical model, with $k_{x,y}$ being multiples of $2\pi/L_{x,y}$, where $L_x$ and $L_y$ are the domain sizes along the $x$- and $y$-axes, respectively. In the models discussed above, the smallest admissible horizontal wave number is $k\mini=2\pi/L_x=2\pi/L_y=\tfrac12\pi\kpc^{-1}$. The smallest wave number along the $z$-axis (along which the boundary conditions are non-periodic) is $2\pi/L_z\approx1.8\kpc^{-1}$. In the nonlinear state, the solution is not a perfectly periodic function of $(x,y)$ within the domain, so that the constraint related to the periodic boundary conditions is much less important.

In the reference model Sim6, the most unstable linear mode has $(k_x,k_y,k_z)\approx(1.6,3.1,1.8)\kpc^{-1}$. This mode has the largest admissible wavelengths in the $x$- and $z$-directions and two complete wavelengths fit along the $y$-direction. In order to confirm that the size of the domain does not affect excessively the parameters of the most unstable mode, we explored the linear phase of the instability using lower-resolution simulations with the same parameters as Model Sim6 but with bigger domains in $x$ and $y$. The results are presented in Table~\ref{LaDo}. Figure~\ref{PoSp} shows how the power spectra of the linear fluctuations depend on the horizontal size of the computational domain. 
The power spectra in $x$ (Fig.~\ref{PoSp}a) do not vary much between the cases $L_x=8$ and $20\kpc$ and suggest that a wide range of modes with $k_x\la2\kpc^{-1}$ are excited by the linear instability. The energy at short wavelengths, $k_x\ga2\kpc^{-1}$, is overestimated in the smaller domain with $L_x=4\kpc$ but a broad maximum at $k_x\approx1.6\kpc^{-1}$ occurs for all values of $L_x$. The power spectra in $y$ (Fig.~\ref{PoSp}b) have a prominent maximum at $k_y\approx3\kpc^{-1}$ in all cases, corresponding to a single most unstable mode. The power spectra in $z$ shown in Fig.~\ref{PoSp}c have a strong maximum at the smallest available wave number $k_z\approx3.6\kpc^{-1}$ for the two smaller domain sizes (representing the large-scale vertical variations shown in Fig.~\ref{fig:zvars}) but modes with $k_z\la3\kpc^{-1}$ dominate when $(L_x,L_y)=(20,24)\kpc$: the solution develops a nearly uniform component in $z$. Remarkably, the horizontal size of the domain affects the vertical structure of the solution. The modes at large wave numbers in all three directions that have an approximately power-law spectrum apparently result from the instability since they are sensitive to the domain size but nonlinear effects, however weak at $t\la0.4\Gyr$, may also contribute. We conclude that the domain size of the models presented in Table~\ref{tab:my_label} is sufficient to capture the properties of the most rapidly growing modes.

\subsubsection{The role of dissipation}\label{TRoD}
The only difference of Model Sim7 from the reference Model Sim6 is the higher (by a factor of 10) gas viscosity $\nu$ and magnetic diffusivity $\eta$ (apart form a lower spatial resolution justified by the fact that stronger dissipation suppresses small-scale structures). The magnitudes of the transport coefficients in Model Sim7 are close to their turbulent values in galaxies. The instability growth rates in the two model are compared in Fig.~\ref{fig:convetatestrms}; the growth rate is hardly affected by the change. However, stronger dissipation leads to a noticeably larger magnetic field strength and cosmic ray energy density in the steady state; the effects of dissipation in the nonlinear state of the instability perhaps deserve further analysis.

\section{Conclusions}\label{sec:Conclusions} 

The nonlinear, saturated state of the Parker instability is different from what might be expected. It is very different from its linear modes. The instability is driven by the gradient of the magnetic field energy density along the gravitational acceleration, and it might be expected that the saturated state would have a reduced gradient of the magnetic field strength. However, the models described here behave differently: not only is the magnetic field gradient reduced, but its strength also reduces throughout the system. The instability produces ubiquitous local vertical magnetic fields which facilitate rapid removal of the cosmic rays from the system by their rapid diffusion along the magnetic field.  As a result, the statistically steady state is notable for a small scale height of the gas and very greater scale heights of the magnetic field and cosmic rays. The system settles into a state close to hydrostatic equilibrium where the gas is supported mainly by the thermal pressure. The regular pattern of magnetic loops and the associated variations in the gas and cosmic rays typical of the linear instability evolves into a rather chaotic nonlinear state where such regular features are hardly noticeable.

The thermal, magnetic and cosmic ray energy densities exhibit significant anti-correlation throughout the system reflecting the pressure balance in horizontal planes. Meanwhile, the magnetic and cosmic ray energy densities are weakly correlated closer to the midplane and anti-correlated further away from it. This behaviour is inconsistent with the energy equipartition between cosmic rays and magnetic field at the scales of the instability of order $1\kpc$.

Apart from gas inflow associated with the reduction of the gas scale height, the instability drives a transient systematic outflow for about $0.3\Gyr$ with the mass flux of order $5\times10^{-2}\,\text{M}_\odot\kpc^{-2}\Gyr^{-1}$ while it evolves through a weakly nonlinear regime. The flow is driven by the magnetic force and involves the vertical magnetic field.

We have also tested the simplifying assumptions used in various analytical (and some numerical) models of the Parker instability to conclude that very few -- if any -- of them are confirmed by the simulations. It appears that analytical analyses of the linear instability provide a reasonable qualitative picture but they should be used with caution when quantitative estimates are required. We find that that the instability growth rate scales more accurately with the scale height crossing time based on the effective sound speed \eqref{ceff} rather than the sound or Alfv\'en speeds.

The models discussed here neglect rotation. A more realistic model would  include (differential) rotation and the dynamo action that supports the unstable background state. Their effects will be discussed elsewhere, and we note that rotation can affect the nonlinear state of the system.
\section*{Data Availability}


The raw data for this work were obtained from numerical simulations using the open-source  PENCIL-CODE available at \url{https://github.com/pencil-code/pencil-code.git}). The derived data used for the analysis presented in the paper are available on request from the corresponding author.


\appendix

\section{The background state and deviations}\label{sec:background}

In order to implement the background distributions of the magnetic field and cosmic rays, that can feed the instability without being changed through its linear and nonlinear stages, we formally introduce source terms $\vec{S}_A$ and $S\cra$ in equations \eqref{ind} and \eqref{ecr} for the magnetic vector potential and cosmic rays, respectively:

\begin{align}
\label{eq:induction_tot}
\frac{\partial\vec{A}}{\partial t} &= \vec{U} \times \vec{B} + \eta \nabla
\times \vec{B} + \vec{S}_{A}\,,\\
\label{eq:crfluid_tot}
\frac{\partial \epsilon_{\rm cr}}{\partial t} &=
-\vec \nabla \cdot \left(\epsilon_{\rm cr} \vec{U}\right) -P_{\rm cr} \left( \vec \nabla \cdot \vec{U} \right)  - \nabla \cdot \vec{F} + S\cra\,.
\end{align}
Since the background velocity vanishes, $\vec{U}_0=\vec{0}$, the background
continuity equation is satisfied automatically whereas the background momentum
equation reduces to the magnetohydrostatic equilibrium
equation~\eqref{eq:MHE} where $P_0$ includes the background magnetic pressure
$P_\text{m0}$, such that $\nabla P_\text{m0} = (4\pi)^{-1} (\nabla\times\vec{B}_0)\times\vec{B}_0$.

The background magnetic field $\vec{B}_0$ corresponds to
\begin{equation}
\vec{S}_A=-\eta \nabla \times \vec{B}_0\,.
\end{equation}
In terms of the magnetic field source this corresponds to $\vec{S}_B=\nabla \times \vec{S}_A$ or, equivalently, $\vec{S}_B=\eta \nabla^2 \vec{B}_0$. The desired magnetic field is aligned with the $y$-axis and only varies with $z$, so that $\vec{S}_B=- \widehat{\vec{y}}\,\eta\, \dd^2 B_{y0}/\dd z^2$. The magnetic source term models the dynamo action responsible for maintaining a stationary, azimuthal field in the disc. 

The source term for the energy density of cosmic rays is given by
\begin{equation}
S\cra = \nabla \cdot \vec{F}_0\,,
\end{equation}
with the background cosmic ray energy density flux given by 
$F_{0_i} = -\kappa_{0_{ij}} \, \partial \epsilon_{\rm cr0}/\partial x_j$.
The divergence of $\vec{F}_0$ represents the Fickian diffusion of the background energy density, and the source maintains the background state against the diffusion. For the background states described in Section~\ref{sec:Theoretical_model}, we have $\vec{F}_0=-\widehat{\vec{z}}\,\kappa_\perp\, \dd \epsilon_\text{cr0}/\dd z$, and $S\cra=-\kappa_\perp\,\dd^2 \epsilon_\text{cr0}/\dd z^2$. The background state is then stationary, as required:
\begin{align}\label{eq:crfluid_imp}
\deriv{\epsilon_\text{cr0}}{t} &= -\nabla \cdot \vec{F}_{0} + S\cra = 0\,,\\
\tau\deriv{F_{0_i}}{t}&=-\kappa_{0_{ij}}\deriv{\epsilon_\text{cr0}}{x_j}-F_{0_i}= 0\,.
\end{align}
Note that, in the cosmic ray energy flux perturbation equation~\eqref{eq:crflux_pert}, the term $-(\kappa_{ij} \partial\epsilon\cra/\partial x_j - \kappa_{0_{ij}} \partial\epsilon_\text{cr0}/\partial x_j)$ represents the relevant term for the perturbation: here the background energy flux is subtracted from the total flux. In practice, there is no need to solve equations~\eqref{eq:induction_tot} and \eqref{eq:crfluid_tot} for the background fields, which can be specified explicitly in their desired forms presented in Section~\ref{sec:Theoretical_model} in equations~\eqref{eq:induction_pert}--\eqref{eq:crflux_pert}. The only purpose of the discussion above is to show that the implementation of the background state adopted here is physically and mathematically sound.
The vertical distributions of the background gas density, magnetic field and cosmic rays can be derived from the magnetohydrostatic equilibrium for a fixed gravitational profile. In terms of  $c\eff$ of equation~\eqref{ceff}, with $c\eff=\const$ in this case, the total pressure is given by $P_0 = P_\text{th0} + P_\text{m0} + P\ecri = c\eff^2\rho_0$.
For a chosen gravity profile and in the magnetohydrostatic equilibrium, we have
\begin{equation}
    c\eff^2\oderiv{\rho_0}{z} = -\rho_0 g \,, \qquad
    \ln \rho_0 =-\frac{1}{c\eff^2}\int_0^z g(z)  \,\dd z\,.
\end{equation}
For $g(z)$ of equation~\eqref{KG89grav} this gives
\begin{align}
    \ln\frac{\rho_0}{\rho_0(0)} = \displaystyle\frac{a_1}{c_{\rm eff}^2} \left(z_1-\sqrt{z^2+z_1^2} 
- \displaystyle\frac{a_2z^2}{2\,a_1\,z_2}\right).
\end{align}
The only non-vanishing component of the magnetic vector potential in the background state then follows from equation~\eqref{B0z} as $A_{x0}=\int_0^z B_0\,\dd z$. 


\label{lastpage}
\end{document}